\documentclass[fleqn,usenatbib]{mnras}
\usepackage[export]{adjustbox}
\usepackage{amssymb,bm}
\usepackage{subfigure}
\usepackage{bm} 
\usepackage{graphicx}
\usepackage{graphicx}
\usepackage{subcaption}
\usepackage{float}
\usepackage{amsmath}
\usepackage{url}
\usepackage[usenames,dvipsnames,svgnames,table]{xcolor}
\usepackage{color}
\definecolor{color1}{RGB}{191, 0, 255}

\def \mnras {MNRAS}
\title{Non-ideal MHD simulations of hot Jupiter atmospheres}
\author[Soriano-Guerrero \& al.]{Cl\`audia Soriano-Guerrero$^{1,2}$\thanks{E-mail: soriano@ice.csic.es}, Daniele Vigan\`o$^{1,2,3}$, Rosalba Perna$^{4}$, Albert Elias-López$^{1,2}$, Hayley Beltz$^{5}$
\\
$^1$Institut de Ci\`encies de I'Espai (ICE-CSIC), Campus UAB, Carrer de Can Magrans s/n, 08193 Cerdanyola del Vallès, Spain\\
$^2$Institut d’Estudis Espacials de Catalunya (IEEC), 08860 Castelldefels, Spain\\
$^3$Institut Aplicacions Computationals (IAC3),  Universitat  de  les  Illes  Balears, 07122, Palma  de  Mallorca\\
$^4$Department of Physics and Astronomy, Stony Brook University, Stony Brook, NY 11794-3800, USA\\
$^5$Department of Astronomy, University of Maryland, College Park, MD 20742, USA \\
}

\date{}

\pubyear{2023}

\begin{document}
\label{firstpage}
\pagerange{\pageref{firstpage}--\pageref{lastpage}}

\maketitle

\begin{abstract}
In Hot Jupiters (HJs), atmospherically induced magnetic fields are expected to play an important role in controlling the wind circulation and in determining their inflated radii. Here we perform 1D plane-parallel magnetohydrodynamic (MHD) simulations of HJ atmospheric columns, using the wind and thermodynamic profiles generated by global circulation models of different exo-planets. We quantitatively investigate the effects of magnetic field winding and Ohmic dissipation (previously considered in several works), with the addition of Hall drift and ambipolar diffusion. 
The main effect is the magnetic field winding in the full non-linear regime, with local azimuthal fields reaching maximum values up to ${\cal O}(10^2)$ G at the shear layer (typical pressure $\sim 1$ bar), much stronger than the assumed background field generated in the planetary interior. The associated meridional currents undergo Ohmic dissipation, with local heating efficiencies of at least $\sim$ ${10^{-6}}-10^{-3}$ (considering only these shallow layers). In addition to the dominant winding vs. Ohmic balance, the presence of the Hall and ambipolar terms have a non-negligible contribution in shaping and twisting the induced magnetic field at $p\lesssim 1$ bar; however this effect is only 
apparent
for the hottest planets. Our results, though limited by construction to a plane-parallel approximation of the sub-stellar columns and with a simplified setup that cannot consistently include the magnetic drag on the wind, assess the non-linearity and complexity of the magnetic induction in HJs atmospheres, and call for a self-consistent inclusion of MHD effects in Ohmic dissipation studies and circulation models, beyond the often-assumed perturbative regime.
\end{abstract}


\section{Introduction}\label{sec:intro}

Hot Jupiters (HJs), known for their distinct properties (e.g. \citealt{knutson07, heng15,fortney21}) represent a unique category of exoplanets. These massive gas giants orbit in extremely close proximity to their host stars, with separations $\lesssim$ 0.1~\rm{au} (astronomical units), and thus experience intense irradiation, with equilibrium temperatures up to $\sim 2200$ K, with a few cases even exceeding it. They undergo tidal locking soon after formation, so that their rotational and orbital periods synchronize; therefore, they exhibit substantial temperature disparities between their illuminated dayside and shadowed nightside, which trigger the formation of powerful zonal jets, working to redistribute the heat ( e.g., \citealt{cho08, dobbs08, showman09, heng11, perna12, rauscher12, batygin13, rauscher13, perez13, parmentier13,rogers14b, showman15, kataria15, koll18, beltz22, komacek22, dietrich22}).

A notable characteristic is that a substantial fraction of these hot giants display remarkable inflation, with radii that can reach up to twice the Jupiter radius in some cases, at odds with the expectations of standard planetary cooling models \citep{showman02,wang15}. This inflation, which is evident for equilibrium temperatures $T_{\rm eq}\gtrsim 1000$~\rm{K}, clearly correlates with $T_{\rm eq}$ (e.g. \citealt{laughlin11,thorngren18}). Despite this correlation, the sole influence of irradiation appears to be insufficient to quantitatively account for the significant expansion in radii: the blanketing effect of the shallow radiation-absorbing layers (above a few bars typically) can keep the planet relatively warm, slowing down the shrinking (i.e., cooling), but can only explain the inflation of up to typically $\sim 1.2-1.3~R_j$. Various possible mechanisms have been proposed. They can be broadly classified into two groups, one invoking planetary evolution with a delayed cooling 
(e.g. \citealt{burrows07,chabrier07}), and the other relies on the presence of additional heat sources (\citealt{bodenheimer01,li10,youdin10,ginzburg15,komacek17}), which are effective especially if they operate in the convective interior of the planet.

The observed correlation with $T_{\rm eq}$ favours the latter, and suggests a continuous deposition of a fraction of the irradiation flux
(typically ranging up to a few percent, see e.g. \citealt{komacek17,thorngren18}), which acts to slow down their long-term cooling and may even lead to reinflation \citep{lopez16,komacek20}. Among the possible heating sources, the Ohmic scenario is a popular candidate to explain the inflation. It consists of the dissipation of currents induced by the stretching of magnetic fields due to jets of the uppermost weakly ionized atmospheric layers \citep{liu08, batygin10, batygin11,perna10a,perna10b,ginzburg16,knierim22}. These studies have primarily operated within the confines of the perturbative or Ohmic-dominated (sometimes called linear) regime of induction, where the induced magnetic field is a perturbation to the internally generated one. Such regime is applicable to the colder HJs only \citep{batygin13,dietrich22,soriano23}, as we will show more in detail in this work.

Other related efforts are global circulation models (GCMs) which include the effects of magnetic fields to some extent. This has been done mainly via 3D magnetohydrodynamic (MHD) simulations (\citealt{batygin13, rogers14a, rogers14b,rogers17}), or via estimates of magnetic drag and Ohmic timescales within hydrodynamical models (e.g. \citealt{rauscher13,  beltz22}). A key ingredient in GCMs is indeed the self-regulating mechanism that limits and distorts the otherwise mainly azimuthal, supersonic winds: the magnetic fields induced by the weakly ionized winds can exert enough Lorentz force to limit their velocity and change their pattern. This effect may qualitatively explain the quite sharp decrease in efficiency (Ohmic power-to-irradiation ratio) for increasing $T_{\rm eq}$, needed to quantitatively explain the inflated radii trend with $T_{\rm eq}$ in the HJ population when evolutionary models including internal heating are employed \citep{thorngren18}.

\cite{soriano23} approached the atmospheric magnetic induction from a different but complementary perspective: a quantitative evaluation of MHD induction in the advection-dominated regime (hot temperatures and large induced fields), considering winding and turbulence alone.  This work has extended the purely hydrodynamic study of \cite{ryu18} by performing 3D ideal MHD simulations with a simplified wind profile and an isothermal background, idealizing a narrow atmospheric column within the dayside radiative layers of the upper atmosphere (mbar to 10 \rm{bar}) of ultra-hot Jupiters (uHJs, $T_{\rm eq}\gtrsim 2500$ K). There, we mostly focused on the 3D effects of the turbulence arising from imposed random perturbations and
we further observed in all the simulations the confinement of a strong azimuthal magnetic field resulting from winding, in qualitative agreement with the non-linear regime studied by \cite{dietrich22}.  
A key caveat in \cite{soriano23}, which restricted the applicability of results to uHJs only, was the use of ideal MHD, with numerical resistivity the only limiting factor to the otherwise endless winding.

This work aims to extend the study of the induced steady-state magnetic field due to the winding effect to HJs of any temperature, incorporating non-ideal MHD (magnetohydrodynamic) effects. Specifically, we aim to find solutions to a general induction equation in 1D plane-parallel approximation. We include the advective, Ohmic, Hall, and ambipolar terms. 
We employ the output of GCM simulations by \cite{beltz22}, \cite{Colombe2023} and \cite{rauscher13} for the sub-stellar points of different models: wind, pressure, and temperature profiles. The profiles of conductivity, the electron and ion densities, entering in the non-ideal induction terms, are calculated according to the imposed thermodynamical profiles. We evolve the MHD equations (including forcing terms to maintain the imposed wind and temperature profiles) until the solution is stationary. We then evaluate the induced magnetic field component and the deviations of thermodynamical quantities from the original background profile. This approach, although intrinsically local and not a substitute for GCMs, allows us to infer the main features of the non-ideal atmospheric induction, which can be taken into account for Ohmic dissipation models, for the effective inclusion of magnetic drag in hydrodynamical GCMs, and can serve as a basis for additional 3D turbulence modeling. 

The paper is structured as follows. In section \S\ref{sec:modelsetup} we present the model of the HJ atmospheric column, together with the GCM input profiles, the code, the numerical methods used in the simulations and the initial and boundary conditions. In section \S\ref{sec:results} we show the results of the study of non-ideal induction effects,
with particular focus on a representative case, the model of WASP 76b. 
Finally in  section \S\ref{sec:conclusions} we discuss the conclusions and final remarks of our work.

\section{Numerical model of a Hot Jupiter atmospheric column}\label{sec:modelsetup}

In this study, we simulate the outermost, radiative layers of a HJ atmospheric column. The concept and the code are similar to the 3D turbulent simulations by \cite{soriano23}, where we focused on very simple wind and hydrostatic, isothermal profiles, using ideal MHD. 
Here we use a 1D version, considering only the vertical variations with a plane-parallel approximation, but we expand on our previous investigation in several aspects. As a matter of fact, we release the ideal MHD approximation, incorporating the three important non-ideal terms, and adopt as inputs the thermodynamic and wind profiles from GCM simulations of \cite{beltz22}, \cite{Colombe2023} and \cite{rauscher13}. For all GCM outputs, here we only consider the substellar point, which is arguably the most extreme in terms of local induction, having temperatures higher than $T_{\rm eq}$, and, therefore, higher conductivity and induced local currents.   
Each ingredient of the model is described in detail in the following.

\subsection{Equation of state and perturbative compressible MHD approach}
\label{subsec:EOS}

We adopt an ideal gas equation of state given by:
\begin{equation}\label{eq:eos}
p = \rho R T = (\gamma - 1)\rho e~,
\end{equation}
where $p$ represents the pressure, $\rho$ the mass density, $e$ the specific internal energy, $T$ the temperature, $\gamma$ the adiabatic index (held constant at 1.4 for this study), $R=k_b/\bar m = (8.254/\mu)\times 10^3$ J/(kg\, K) the specific gas constant, where $k_b$ is the Boltzmann constant, $\bar m=\mu m_u$ the mean relative molecular mass of the gas, and $m_u$ the atomic mass unit. 
For simplicity, here we consider a constant value $\mu=2$, thus neglecting variations in chemical composition and related phenomena like cloud formation.
Under the presence of gravity directed downwards in the vertical ($z$) direction, ${\bf g}=-g \hat{z}$, the equation of hydrostatic equilibrium reads
\begin{equation}\label{eq:hydro_eq}
\frac{dp}{dz} = -\rho g = -\frac{pg}{RT} ~.
\end{equation}
We assume that the gravity of a given planet is constant
as a function of height within the atmosphere,
and given by the planetary mass and radius (gravity can vary by maximum $\sim 10\%$ along the range of pressures in the considered columns). This hydrostatic equilibrium is solved, for a given $p(T)$ profile (see \S\ref{sec:gcm}) to set the vertical profiles $p_0(z)$ and $\rho_0(z)$ as a background state.

With these background profiles, we solve the compressible MHD equations in a column, in Cartesian coordinates (like in \citealt{soriano23}), where the $x,y,z$ components of velocity ($\mathbf{v}$), electric ($\mathbf{E}$) and magnetic ($\mathbf{B}$) fields would correspond to the azimuthal, meridional and radial components, respectively, in spherical coordinates. However, unlike the 3D study by
\citealt{soriano23}, here we consider a 1D problem, with a purely vertical domain $z \in [-D,0]$ (where $D$ is the domain height) and consider only vertical dependences since we are interested in the main magnetic effects caused by the dominant, vertical gradients set by the background state and wind profile. As in \cite{soriano23} and in many e.g. solar atmospheric MHD studies \citep{Lou93,Shen2007,felipe10,Zhao2021} which cover many scale heights ($\sim 10$ in our case), we use a perturbative approach, i.e. we evolve the perturbed field, in particular the density perturbation $\rho_1 := \rho - \rho_0$, instead of the total density $\rho$. 

Practically, we set up a 1D, plane-parallel problem and evolve the deviations of density $\rho_1$, the components of the momentum density, $\mathbf{S}=\rho\mathbf{v}$ of the magnetic field, and the total energy density U, which includes internal, kinetic, and magnetic energy contributions. Keeping in mind that  the gradient, divergence and curl operators essentially consist of non-zero $\partial_z$ derivatives only, the continuity, momentum, induction and energy equations, in conservative form, read as follow:

\begin{eqnarray}
 \frac{\partial \rho_1}{\partial t} & + &\nabla \cdot (\rho \textbf{v}) = 0\,, \label{eq:MHDcontinuity} \\
 \frac{\partial \textbf{S}}{\partial t} & + &\nabla\cdot\left[\rho\mathbf{v v}+ \left(p + \frac{B^2}{2\mu_{0}}\right)\textbf{I}-\frac{\textbf{B}\textbf{B}}{\mu_{0}}\right] = \rho \mathbf{g} + \mathbf{F},\label{eq:momentum} \\
 \frac{\partial \mathbf{B}}{\partial t} & = & -\nabla \times \mathbf{E}\,,\label{eq:induction} \\ 
 \frac{\partial U}{\partial t} & + & \nabla \cdot\left[ \left(U + p + \frac{B^2}{2\mu_{0}}\right)\textbf{v} - (\textbf{v} \cdot \textbf{B})\frac{\textbf{B}}{\mu_{0}}\right] = \nonumber \\ 
 && = \rho \textbf{v} \cdot \textbf{g} + \textbf{v} \cdot \mathbf{F}-\frac{p_{1}}{\tau_{cool}}  + Q_j 
 ,\label{eq:MHDenergy}
\end{eqnarray}
where $\textbf{I}$ is the identity diadic tensor, $Q_j$ is the Ohmic heating rate per unit volume (see below for its definition), and $\tau_{cool}$ is the Newtonian cooling timescale (commonly used also in GCM, e.g. \citealt{batygin13,rogers14a,rogers14b,rogers17}), which is used here as an effective way to stick reasonably close to the background profile $T$ and avoid a continuous increase of the internal energy due to numerical and physical dissipation.\footnote{Since we stick to background profiles $p(T)$ computed with 3D GCM model, and we aim at studying the dominant 1D magnetic effects for these profiles, we do not include a more consistent heat diffusion scheme.}

The forcing term in the momentum equation is defined as $\mathbf{F}:=(\rho_{0}F_{0}(v_{x}-v_w(z)),0,-\sigma_{z}(z)S_{z})$. Its $x$-component allows the azimuthal velocity, $v_x$, to be kept very close to the wind imposed profile, $v_w(z)$. The vertical component of the forcing instead includes a damping term which is non-zero only for the uppermost region, $z>z_d$: $\sigma_{z}=A_d\left(\frac{z_d-z}{z_d}\right)^2$. It is numerically needed to maintain hydrostatic stability when the domain covers many scale heights. For a more detailed discussion on this, and on the values to which we fix the forcing parameters ($\tau_{cool}=10$, $F_0=1$, $A_d=10$, $z_d=-0.2\,D$), we refer the reader to App. A of \cite{soriano23}.

By using our hybrid approach, i.e. a full MHD system of equations plus a forcing on the vertical profiles of $p$ (i.e., $T$ and $\rho$) and $v_x$, we can partially evaluate the vertical distribution of density/pressure perturbations and the magnetic feedback on the vertical and meridional fluid velocity, which in turn enters in the induction equation (see below). In general, the thermodynamic deviations are much smaller than the background values, being non-negligible only in some particular ranges, usually in correspondence of low density or high shear (i.e., high induced field). Therefore, although our local box simulations with given input profiles cannot consistently include the global feedback of magnetic effects on the flow and on the $p(T)$ profiles, we can roughly assess at which depths and by how much the induced magnetic fields are expected to change the considered thermodynamic profiles, if they were fully included in GCM models.

\subsection{Induction equation and conductivity profile}
\label{subsec:MHD}

The simulations presented in this work focus on the upper regions of a hot Jupiter atmosphere, with typical pressure and temperature ranges of $p \sim 10^{-3}$-$10^2$ bar,  $T \sim 1000{-}3000$~K. In this region, thermal ionization of alkali metals occurs \citep{parmentier18, kumar21, dietrich22}, while pressure ionization of hydrogen is relevant and dominant only in much deeper layers, $p\gtrsim 10^6$ bar \citep{French12,bonitz24}. As a result, the interplay of strong zonal winds and a magnetic field generated in the planetary interior induces the circulation of electrical currents within the atmosphere. In turn, these currents can generate a strong magnetic field in the direction of the wind, via advection. Besides the advective and Ohmic terms, other effects, non-linear in {\bf B}, may play a minor role at the considered pressure: the Hall drift and ambipolar diffusion. With these premises, applicable to a weakly ionized medium, the electric field entering in the induction equation reads (e.g. \citealt{pandey08,koskinen14}):
\begin{equation}
 \mathbf{E} = - \mathbf{v} \times \mathbf{B} + \frac{\mathbf{J}}{\sigma} + \frac{\mathbf{J} \times \mathbf{B}}{e n_{e}} - \frac{\left( \mathbf{J} \times \mathbf{B} \right) \times \mathbf{B}}{ \nu_{in} \rho_i}\, , \label{eq:electric_field}
\end{equation}
where: $\textbf{J}$=($\nabla\times\textbf{B}$)/$\mu_{0}$ is the electric current density, $\mu_0$ the magnetic permeability, $\sigma$ is the local electric conductivity, $e$ the elementary charge, $n_{e}$ the number density of electrons, the most relevant charge carriers, $\rho_i$ the mass density of ions, $\nu_{in}$ the collision rate between ions and neutrals, which, for an ideal, classical gas, is given by \citep{pandey08,draine83}:
\begin{equation}
\nu_{in}=\rho_n\frac{\langle\sigma_c v_{in}\rangle_i}{\bar m_{i}+\bar m_n} \simeq \rho\frac{\langle\sigma_c v_{in}\rangle_i}{\bar m_{i}+\mu m_u}~,\label{eq:nu_in}
\end{equation}
where $\langle\sigma_c v_{in}\rangle_{in}=1.9\times10^{-9}$~cm$^3$~s$^{-1}$ is the collision rate coefficient (average of the product between the cross-section $\sigma_c$ and the relative velocity $v_{in}$) between ions and neutrals, $\bar m_{i}$ is the mean ion mass, for which we adopt the value $\bar m_{i}=30~m_{p}$, which is a representative value for the expectedly most abundant ions,
K$^+$ and Na$^+$.

The electric field in eq.~(\ref{eq:electric_field}) consists of the ideal term (wind advection in our scenario), resistivity, Hall effect, and ambipolar diffusion, respectively. We have neglected the terms proportional to the pressure gradient, which would give the Biermann battery effect and contribute further to the ambipolar diffusion. Such terms are safely negligible, since in our case the \'Alfven speed is much smaller than the sound speed \citep{pandey08}. Note also that we are considering pressures $p\gtrsim$ mbar, for which eq.~(\ref{eq:induction}) applies, since the collision frequencies of both ions and electrons are larger than the electron plasma frequency. This implies that charged components are strongly coupled to the neutral one, so that we can neglect electron plasma waves and the anisotropic nature of the conductivity (we are in the M1 region of \citealt{koskinen14}).

{Since we consider only vertical gradients, the only relevant electrical components are $E_x$  and $E_y$, which are responsible of inducing the azimuthal ($B_x$) and meridional ($B_y)$ magnetic field.
Overall, at first order, the dominant contributions are expected to come from the winding mechanism (advective term, $\partial_z(v_xB_z)$), and Ohmic dissipation, $\partial_z(J_y/\sigma)$, especially for deep enough layers. 
Depending on the relative weight of the advective to resistive terms, the induced field can be just a perturbation of the background field (Ohmic-dominated regime), or be locally comparable to or even larger than it (advection-dominated regime), see \cite{dietrich22} for an in-depth discussion. Note that the winding mechanism is intrinsically linear: the induced field grows linearly with the component of the magnetic field which is perpendicular to the flow. However, the advection-dominated regime is sometimes called non-linear, in the sense that, in a full 3D GCM, the induced field can have a non-trivial feedback on the flow, which in turn modifies the direction and intensity of the induced field, so that considering the background field alone in the $(\mathbf{v}\times\mathbf{B})$ term is not valid anymore (see e.g. the analytical and numerical study by \citealt{batygin13}), unless one considers a very idealized case (axial symmetry, purely azimuthal wind, alignment between rotational axis and magnetic moment).

 Compared to these two dominant terms, Hall and ambipolar effects are expected to be negligible in dense environments, but they could be relevant for the uppermost layers (\citealt{perna10a}), especially if the locally induced magnetic field is large enough, since their relative importance with respect to the Ohmic term scales as $B$ and $B^2$, respectively. Qualitatively, the Hall drift tends to introduce a drift orthogonal to the magnetic field, without causing any direct dissipation, while the ambipolar diffusion arises from a non-zero velocity difference between ions and electrons, which can be evaluated from  eqs.~(\ref{eq:induction}) and (\ref{eq:nu_in}):
\begin{equation}
v_i - v_n = \frac{\textbf{J} \times \textbf{B}}{\rho_i \nu_{in}}~. \label{eq:v_amb}
\end{equation}
The ambipolar term is expected to play an important role in tenuous partially ionized plasma, especially in the widely studied case of the solar photosphere and its overlying layers (e.g., \citealt{chitre01,popescu21}). If we rewrite the ambipolar term as:
\begin{equation}
-\frac{1}{\nu_{in}\rho_i}\left(\textbf{J} \times \textbf{B} \right)\times \textbf{B} = \frac{1}{\nu_{in}\rho_i}[B^2\textbf{J} - (\textbf{J} \cdot \textbf{B})\textbf{B}]\, 
\end{equation}
we can see that part of the ambipolar term (the first term in the right-hand side) tends to dissipate part of the currents, to have them aligned to the magnetic field (i.e., a so-called force-free configuration, with no Lorentz force). In HJs, indeed, the winding term creates a configuration with an induced current $\textbf{J}\perp\textbf{B}$, i.e., a Lorentz force which, in the simplest axisymmetric, aligned case, go exactly against the wind direction \citep{batygin13}. Such a term enters in the ambipolar term in our local simulations, and represents a large drag term in the momentum equations in GCMs. Moreover, since the ambipolar effect scales with $B^2$, regions with larger winding-induced field will undergo ambipolar dissipation more easily.

The electrical conductivity, which regulates the Ohmic term, is calculated locally along the whole vertical column according to the background $p(T)$, considering the classical formula, e.g. \cite{draine83}:
\begin{equation}
  \sigma(T,p) =\frac{x_e}{\langle\sigma v_{en}\rangle_e} \frac{e^2}{m_e}\,,
\label{eq:sigma}
\end{equation}
where $m_e$ is the electron mass, $\langle\sigma_c v_{en}\rangle_e$ is the momentum transfer rate coefficient between electrons and neutrals (the dominant channel), which depends on the temperature \citep{draine83},
\begin{equation}
\langle\sigma_c v_{en}\rangle_e (T)  = 10^{-19} \left(\frac{128 k_{\rm B} T}{9 \pi m_e}\right)^{1/2} {\rm m}^3\,{\rm s}^{-1}\,,
\end{equation}
and, for simplicity, we only consider the potassium ionization in the electron fraction $x_e$:
\begin{eqnarray}
 x_e(T,p)  &=&  7.7 \times 10^{-4} \left(\frac{a_{K}}{10^{-7}} \right)^{1/2} T_{2000}^{3/4} \nonumber\\
 && \times \left(\frac{n_{10}}{n_n(T,p)}\right)^{1/2} e^{-\alpha/T_{2000}}~,
 \label{eq8}
\label{eq9}\end{eqnarray}
where $n_n \simeq \rho/(\mu m_u)$ is the neutral number density, which takes the value $n_{10}=3.62\times 10^{25}$ m$^{-3}$ at $p=10$ bar and $T=2000$ K, $T_{2000}$=$T/2000$ K, $\alpha = 12.594$ is a numerical factor, and $a_{K}$ is the potassium mass fraction  ($\sim 10^{-7}$ in the Sun). 
Eq.~(\ref{eq:sigma}) is an analytical approximation to the Saha equation, strictly valid if there are no other ionized elements, and if $x_e$ itself is less than the element mass fraction $a_K$. At high temperatures $T\gtrsim 2500$ K, this approximation becomes less reliable; however, we checked that the results of \cite{kumar21}, who considered the conductivity from solving the full Saha equation for a more diverse chemical composition, provides results in good agreement with the approximation above, with a maximum deviation of a factor of a few in regions with largest $T$. Given that the conductivity varies by orders of magnitude, we consider the potassium-only analytical approximation to be good enough for our purposes.

Finally, the Ohmic dissipation per unit volume, entering in eq. (\ref{eq:MHDenergy}), is defined as:
\begin{equation}
Q_{j}=\frac{J^2}{\sigma}\,,
\label{eq:qj}
\end{equation}
which can be integrated along our column of size $D$, and compared with the flux irradiating it, $F_{\rm irr}=4\sigma_{\rm sb}T_{\rm eq}^4$ (where $\sigma_{\rm sb}$ is the Stefan-Boltzmann constant), to obtain the local heating efficiency
\begin{equation}
\epsilon = \frac{\int_D Q_j(z) dz}{F_{\rm irr}}~.
\label{eq:efficiency}
\end{equation}
Note that this is a local definition (fluxes are defined per unit surface) apt for our local simulation, while normally the heating efficiency is defined at a global level, i.e., integrating over the entire planet and considering deeper, convective layers (well outside our numerical domain), where heat deposition has a pronounced impact on inflation (e.g., \citealt{batygin10,wu13,ginzburg15,komacek17,thorngren18}).

\subsection{Input from GCM models}\label{sec:gcm}

In order to prescribe a column profile for the background ($p_0(z)$ and $\rho_0(z)$) and the wind ($v_w(z)$), we make use of profiles calculated from GCM models of specific planets. The GCM solves a set of fluid dynamical equations known as the ``primitive equations of meteorology'' to simulate a planet's atmosphere. We consider profiles at the sub-stellar point of each setup, covering a significant fraction of the hot Jupiter parameter space, described in Table \ref{table1}. A summary of these inputs can be found in Fig.~\ref{fig:wind,pressure}, where we show the $p(T)$ and $v_w(z)$ profiles for the different cases considered.

Such profiles are output from previous works, including \cite{Colombe2023} for the WASP 18b model, and \cite{beltz22} for the models of WASP 76b. The models for HD 209458b and HD 189733b are equivalent to those published in \cite{rauscher13}. Each model was calculated at a resolution of T32 (corresponding to roughly 3 degrees separation at the equator) and ran until a steady state was reached, corresponding to at least 1000 planetary orbits. 
Those works make use of the Rauscher and Menou (RM)-GCM\footnote{\url{https://github.com/emily-rauscher/RM-GCM}} \citep{rauscher12}, incorporating an updated radiative transfer scheme from \cite{Roman17}, which is based on \cite{toon89}. The set of models used in this paper fall under two categories for radiative transfer: double gray and correlated-K. The models for HD 209458b and HD 189733b use a double-gray approach, employing two absorption coefficients: one for the visible wavelength range to account for the absorption of the host star's radiation, and another for the infrared range to represent the planet's thermal emission; for further details see \cite{rauscher12}. The remaining GCM models use a ``picket fence'' radiative transfer scheme, which uses 5 absorption bands compared to the two used by the double-gray method. For more details on the picket fence radiative transfer, see \cite{Malsky2024}.  

Additionally, all the GCM input models, except WASP 76b-d0, make use of a spatially varying drag timescale that approximates the Lorentz force felt by charged atmospheric species. The background magnetic field strength, $B_d$ is shown in the last column of Table \ref{table1}. The inclusion of this active drag reduces circulation efficiencies and increases the day-night contrast. This magnetic drag is calculated based on local atmospheric conditions and applied in a geometrically and energetically consistent way 
using the expression
\citep{perna10a}:
\begin{equation}
    \tau_{\rm drag}\sim \frac{\rho c}{B_d^{2}\sigma |\cos(\theta)|}\,,
\end{equation}
where $\theta$ is the latitude, and the background field $B_d$ is a free parameter of the model, taken as constant in depth. This expression is a good approximation of the drag timescales if: (i) the induced field is only a perturbation of the background field; (ii) the magnetic field is approximated as a pure dipole (which gives the $\cos\theta$ dependence), aligned with the axis of rotation; (iii) the background field strength does not vary much as a function of radius.

With these assumptions, at the sub-stellar point, which lies on the equator, the magnetic field is expected to be purely meridional, with a zero radial component. However, misaligned magnetic fields, composed by a combination of multipoles, are seen both observationally from e.g. the Jovian magnetic spectrum inferred directly from Juno low-orbit measurements \citep{connerney22}, and theoretically through dynamo simulations of gas giants, e.g. the state-of-the-art simulations by  \cite{gastine21}. Both features (misalignment and non-trivial multipolarity) would result in a non-zero radial component of the magnetic field at the equator. Future GCM works will explore the effects of tilted dipoles or more complex geometries. For this reason, we introduce a small but non-zero initial radial component, that we fix by default as $B_z^{\rm in}= 0.1\,B_y^{\rm in}$ (if not indicated otherwise), which acts as a necessary seed to trigger the winding mechanism in our 1D setup. See also \S \ref{sec:bc} for boundary conditions. Moreover, as part of this paper, we will examine how the induced field likely contrasts with the typically made assumptions.

In Fig.~\ref{fig:wind,pressure} we plot the profiles employed in this study, which span a range of pressures of $\sim$ 0.005-100 bars for all models. A wide range of temperatures $\sim$ 1500-3500 K can be observed (top panel), depending on the pressure and the planet. In particular, for those models with higher equilibrium temperatures, there is a characteristic temperature inversion region where temperature decreases as $p$ decreases. This feature only occurs in the hottest models:
WASP 76b, WASP 76b-d0, WASP 18b, and WASP 121b, 
and is not found in the models for HD 20958b or HD 189733b.

In the bottom panel of Fig. \ref{fig:wind,pressure}, we show the profiles of the azimuthal (east-west) wind. The meridional components (north-south) of the winds are 1-2 orders of magnitude lower than the dominant east-west winds. This is primarily due to the influence of thermal tides induced by stellar radiation, which generate a super-rotational eastward jet at the equator \citep{Menou19, showman02, dobbs08, Showman2008}.

Among the modelled planets, WASP 18b-20 is the one with the highest equilibrium temperature and gravity. The high gravity of this planet results in a much thinner domain (see $D$ values in Table \ref{table1}), compared to the other models. Moreover, this model
also used the strongest background magnetic field strength, $B_d = 20$~G (chosen based on fitting the \textit{JWST} light curve of the planet, \citealt{Colombe2023}) causing the resulting east-west winds to be slower for a larger faction of the modeled domain than in other planets. For the other planets, the GCMs  adopted instead a value of $B_d=3$~G. For comparison, we also consider the profile for a WASP 76b GCM model without magnetic drag, $B_d=0$~G.

\begin{table}
\centering
\begin{tabular}{|c|c|c|c|c|c|c|}
\hline
Model & $M_p$ & $R_p$ & $T_{\rm eq}$ & $F_{\rm irr}$ & $B_{\rm d} $\\
& $[M_{J}]$& $[R_{J}]$ & [K] & [MW/m$^{2}$] & [G] \\
\hline  WASP 76b-d0 & 0.92 & 1.83 &  2160 & 4.9 & 0\\
  WASP 76b & 0.92 & 1.83 & 2160 & 4.9 & 3\\
 HD 189733b & 1.13 & 1.13 & 1191 & 0.32 & 3\\
 HD 209458b & 0.73 & 1.36& 1484 & 1.1 & 3 \\
 WASP 18b & 10.2 & 1.24 & 2413 & 7.7 & 20\\
 WASP 121b & 1.16 & 1.75 & 2358 & 7.0 & 3\\
\hline
\end{tabular}
\caption{List of the main properties of the input profiles used in this work. For each planet modelled by the GCM, labelled as in the first column, we indicate: the observed values (retrieved from the NASA Exoplanet Archive) of mass $M_p$, radius $R_p$ (in units of Jupiter mass and radius, respectively), equilibrium temperature $T_{\rm eq}$ and irradiation $F_{\rm irr}$, and the background magnetic field value $B_d$ used in the drag term in the GCM.}
\label{table1}
\end{table}

\begin{figure}
\centering

\includegraphics[width=.95\linewidth]{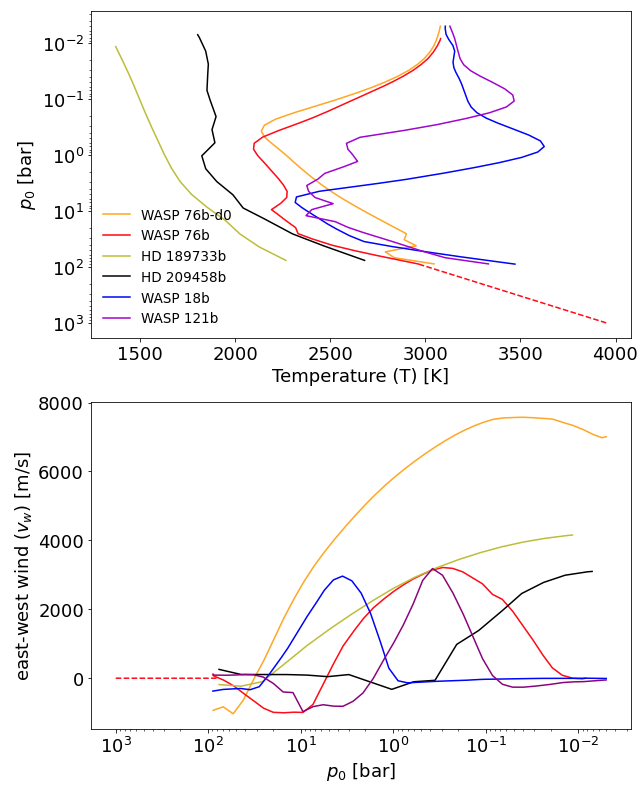}\\
\caption{Substellar profiles for the different models shown in Table \ref{table1}: $p(T)$ (top) and zonal winds $v_w(z)$ (bottom). We show the extension to deeper pressures for WASP 76b with a dashed line.}
\label{fig:wind,pressure}
\end{figure}

Using the $p(T)$ profiles from the GCM models, we recover the hydrostatic equilibrium profiles ($p_0(z),\rho_0(z),T(z)$) by solving eq.~(\ref{eq:hydro_eq}) for the different cases.\footnote{In order to minimize the discretization errors when solving the hydrostatic solution, we have interpolated on a much finer grid the GCM output $p(T)$, which has typically only 30-100 points.} The $p(z)$ and $T(z)$ profiles are shown in the top and middle panels of Fig. \ref{fig:Tandsigma}. There, $z=0$ corresponds to the values in the range $p_{\rm top} \in (5\times10^{-3},10^{-2})$, depending on the simulation. Using these profiles, we compute the conductivity, $\sigma$, based on eq.~(\ref{eq:sigma}) and shown in the bottom panel of Fig. \ref{fig:Tandsigma} as a function of pressure $p_0$. Planets with higher equilibrium temperatures display higher conductivity over the modelled domain. Moreover, if we carefully analyze individual profiles, especially those with higher temperatures, we can appreciate that higher conductivity values are reached near the top of the domain as both the higher temperatures and the lower pressures enhance the conductivity (see eq.~\ref{eq:sigma}). At deeper layers, we encounter regions of temperature inversions (decreasing temperature) followed by a region of increasing temperature. Correspondingly the conductivity, after a decline, increases again, albeit to a lesser extent due to the higher density of the deeper levels. 
HD 1898733b and HD 209458b, being colder than the other planets, have much lower conductivities. The former in particular do not show temperature inversion.

\subsubsection{Extension of the WASP 76b model to deeper layers}\label{sec:extension}

In order to test the influence of considering deeper layers in our simulations we extend the  temperature, wind, and conductivity profiles of our representative case, WASP 76b. This is displayed in Fig.~\ref{fig:wind,pressure} and \ref{fig:Tandsigma} with dashed lines. We extend the model up to 1000 bar, i.e. connecting to the convective region, which, for a Hot Jupiter, can be between $\sim 10^2$ and $10^3$ bar, depending on the amount of irradiation and internal heating \citep{komacek17,thorngren19}. Specifically, we impose an adiabat $p \propto T^\frac{\gamma-1}{\gamma}$, with a euristically chosen index $\gamma=1.12$ which enables us to smoothly connects with the GCM $p(T)$ profile (the precise value of the slope is a second-order effect, for our general purposes).

Regarding the extrapolation of the wind profile, we note that in regions of higher pressure, $p\gtrsim$ 10 bar, the wind speeds tend to greatly decrease (Fig.~\ref{fig:wind,pressure}). As a matter of fact, the azimuthal temperature gradients present in these layers can still drive significant winds, but the increased density and pressure tend to drastically moderate the speeds. For this reason, we assume $v_w$=0 for the extended range of pressure, p $\gtrsim$ 100 bar.

\begin{figure}
\centering
\includegraphics[width=.95\linewidth]{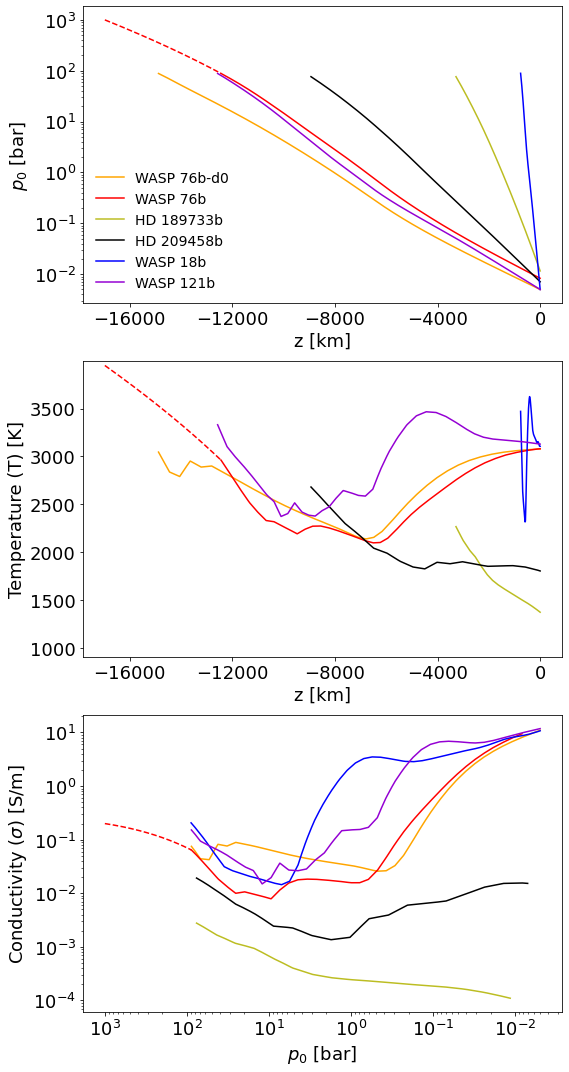}
\caption{Vertical profiles at hydrostatic equilibrium, for the different models shown in Table \ref{table1}, of: pressure $p_0(z)$ (top), temperature $T(z)$ (center) and the corresponding electrical conductivity $\sigma(p)$ (bottom). The layer $z=0$ corresponds to the top boundary of the numerical domain, $p_{\rm top}$.} 
\label{fig:Tandsigma}
\end{figure}

\subsection{Infrastructure and numerical setup}
\label{sec:infra}

In order to perform our simulations we use v3.7.3 of Simflowny \citep{arbona13,arbona18,palenzuela21}, a user-friendly platform that generates codes for partial differential equations, allowing a variety of numerical schemes in finite difference and finite volumes. It employs the SAMRAI\footnote{\url{https://computing.llnl.gov/projects/samrai}} infrastructure \citep{hornung02,gunney16} for the management of the parallelization, mesh refinement (here not used), and output writing. We employ the high-resolution shock-capturing method MP5 for the spatial discretization, using a splitting flux scheme, and a Runge-Kutta fourth-order scheme for the time advance \citep{palenzuela18}.

The bottom domain is characterized by a pressure ranging from 76.44 to 1000 \rm{bar}. The pressure at the top boundary is the minimum value which differ from model to model but lies in the range $p_{\rm top}\sim$ 0.002-0.02.  The upper fifth of the domain, in which we artificially damp the vertical momentum for stability purposes (see \S\ref{subsec:EOS}),  corresponds to pressures $p\lesssim 0.01-0.09$ bar, with the specific value depending on the planet.
We employ a discretization of $N_{z}$=400 grid points in all the simulations presented, except where specified otherwise. This resolution has been chosen to balance computational efficiency with the need of numerical convergence of the solution, which may be demanding in the uppermost, highly conductive layers of the hottest models. Said in another way, the chosen resolution is enough to ensure
that the physical resistivity is higher than the numerical one almost everywhere in all cases, see App.~\ref{appendixA}. 

\subsection{Initial and boundary conditions}\label{sec:bc}

As initial conditions, we impose the hydrostatic profiles $p=p_0(z)$, $\rho=\rho_{0}(z)$, $v_x=v_w(z)$ for each model, as discussed above. The magnetic field is initially set as a purely homogeneous poloidal field, parametrized by constant values $B_y^{\rm in}$ and $B_z^{\rm in}$, representing the meridional and radial components of the background field at the equator, respectively. The azimuthal component is initially vanishing, $B_x^{\rm in}=0$. By default, we use the same value of the background field used in the GCM drag timescale, $B_y^{\rm in}=B_d$, and a smaller value, $B_z^{\rm in}=B_d/10$, for the vertical component (which is the seed for the winding mechanism, hence cannot be exactly zero). However, we also explore different values of $B_y^{\rm in}$ and $B_z^{\rm in}$ for WASP 76b, which we use a reference model. We list in Table \ref{table2} the different setups shown in the paper.

At both boundaries, we impose flat boundary conditions, i.e. a zero $z$-derivative of $\rho_1$, $p_1$, $v_x$ and $v_y$ (meaning, for the latter two, a stress-free boundary condition), from which we derive the values of $S_x,S_y,\textbf{U}$ by using their definition. For consistency with the continuity equation at equilibrium, we impose $\partial_zS_{z}=0$,\footnote{The presence of a non-zero $v_z$ and the necessity of a damping term in the uppermost layers don't allow a perfectly constant $S_z$ at equilibrium: however, if we remove the damping term, the velocity grows uncontrollably, for large enough domains \citep{soriano23}.} and consequently $v_{z}=S_{z}/\rho$.
For the magnetic field, we fix all components $B_x$, $B_y$ and $B_z$ to their background, initial values, both at the top and the bottom boundaries. In other words, we are confining the atmospheric induction only to the considered domain. This is a conservative choice and neglects the extension of induced currents at deeper layers \citep{liu08,batygin10,wu13,ginzburg16,knierim22}. This assumption is in part dictated by the need to have a numerically convergent solution (see App. \ref{appendixA}). We will discuss the caveats and implications of this assumption in \S \ref{subsec:depth}, showing how the results change when we consider deeper layers, for the  extended setup of WASP 76b explained in \S \ref{sec:extension}.

\section{Results}
\label{sec:results} 

\begin{table*}
\centering
\begin{tabular}{|c|c|c|c|c|c|c|c|c|}
\hline 
Model & $B_{z}^{\rm in}$ & $B_{y}^{\rm in}$ &$p_{\rm top}$ & $p_{\rm bot}$ & $D$ & $|B_{x}|_{\rm max}$ & $\int_D Q_{j} dz$ & $\epsilon$  \\
& [G] & [G] & [bar] & [bar] & [km] & [G] & [MW/m$^2$] & \\
\hline 
WASP 76b & 0.3 &3 & 0.002 & 88.34 & 18762 & 180 & $8.0\times10^{-4}$ & $1.6\times10^{-4}$ \\ 
HD 189733b & 0.3 & 3 & 0.02 & 76.44 & 2830 &0.39 & $2.9\times10^{-6}$ & $9.1\times10^{-6}$ \\
HD 209458b & 0.3 & 3 & 0.007 & 76.44 & 8190 & 14 & $8.5\times10^{-5}$ & $7.7\times10^{-5}$  \\
WASP 18b& 2&20 & 0.004 & 89.13 & 770 & 260 & $7.5\times10^{-3}$ & $9.7\times10^{-4}$ \\
WASP 121b & 0.3 & 3 & 0.003 & 88.34 & 13230 & 1550 &$8.0\times10^{-3}$ & $1.1\times10^{-3}$\\ 
\hline 
WASP 76b-p300$^{*}$ & 0.3 & 3 & 0.005 & 300 & 19249 & 210 & $8.1\times 10^{-
4}$ & $1.7\times10^{-4}$ \\
WASP 76b-p500$^{*}$ & 0.3& 3 & 0.005 & 500 & 20367 & 275 & $7.8\times 10^{-4}$  & $1.6\times10^{-4}$ \\
WASP 76b-p1000$^{*}$ & 0.3 & 3 & 0.005 & 1000 & 21979& 390 & $7.8\times 10^{-4}$ & $1.6\times10^{-4}$ \\
\hline

WASP 76b-d0-Bz0.3$^{*}$ & 0.3 & 0 &0.005&88.34 & 17220 & 870 & $3.5\times10^{-3}$ & $7.1\times10^{-4}$ \\ 
WASP 76b-d0-Bz0.03$^{*}$ & 0.03 & 0 & 0.005 & 88.34 & 17220 & 80 & $3.1\times10^{-5}$ & $6.3\times10^{-6}$ \\
WASP 76b-d0-Bz0.003$^{*}$ & 0.003 & 0 & 0.005 & 88.34 & 17220 & 8 & $3.1\times10^{-7}$ & $6.3\times10^{-8}$ \\
\hline
\end{tabular}
\caption{Input properties of the models considered in this work. In each column, we indicate: the model, the initial value of radial ($B_z^{\rm in}$) and meridional ($B_y^{\rm in}$, taken equal to the background field $B_d$ assumed in the GCM models on which we base our profiles), the pressure at the top and bottom of the domain we consider the corresponding size $D$ of the numerical domain, the maximum absolute value of the main induced component $B_x$, the integrated Ohmic dissipation rate, and the corresponding heating efficiency $\epsilon$, eq.~(\ref{eq:efficiency}). The asterisk indicates purely winding+Ohmic cases (i.e., without Hall and ambipolar terms), for which the solution is completely independent on $B_{y}^{\rm in}$.}
\label{table2}
\end{table*}

\subsection{General behaviour}\label{sec:resultsintro} 

We let the system evolve until all terms in the electric fields find a balance and a stationary solution is reached, i.e., spatially uniform electric field components $E_x$, $E_y$. In all the simulations, at the beginning, the main effect is the creation of the initially inexistent azimuthal field, $B_x$, via winding, as long as there is a non-zero vertical component ($B_z$). At first order, the induced field grows linearly until the Ohmic term associated with the induced meridional currents, $J_y$, grows enough to limit it:
\begin{equation}
    \frac{\partial B_{x}}{\partial t} \simeq \frac{\partial v_{x}}{\partial z} B_{z}+\frac{1}{\sigma}\frac{\partial J_{y}}{\partial z} \simeq 0 ~.
\label{eq:induction2}
\end{equation}
The resulting magnetic field lines are highly wound up. At the same time, the $B_y$ component can deviate from the initial, background value only due to the Hall and ambipolar terms, which in turn highly depend on the winding-induced component $B_x$ and its associated current component $J_y$. Note that there is no induced $B_z$, by construction, in our 1D approach.

\begin{figure}
\centering
\includegraphics[width=.9\linewidth]{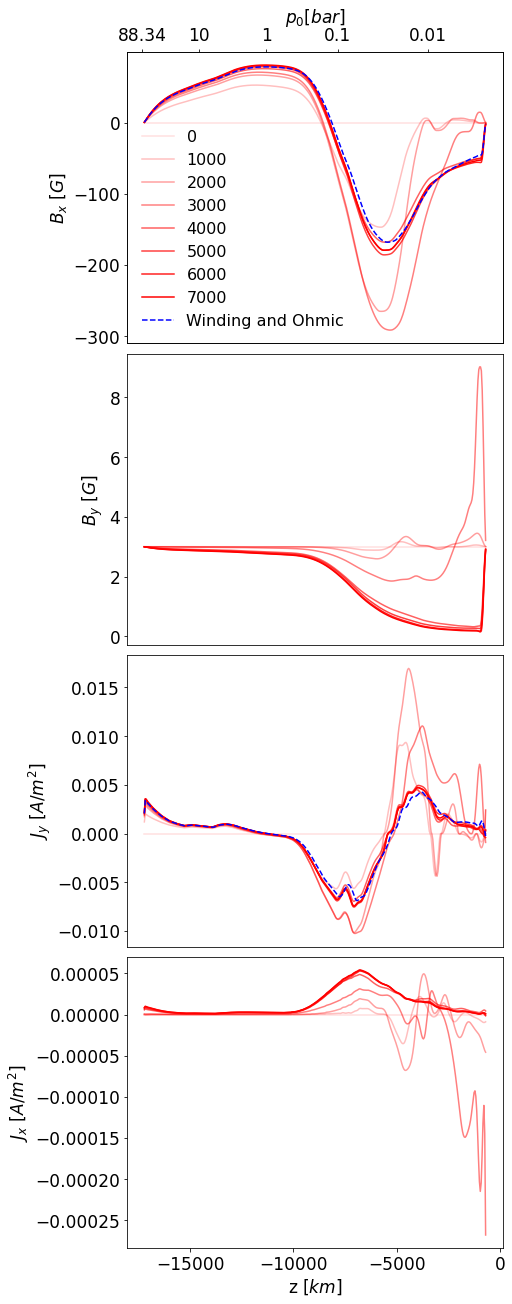}\\
\caption{The vertical profiles of $ B_{x}$(z) (top), $B_{y}$(z) (upper center), $J_{y}$(z) (lower center) and $J_{x}$(z) (bottom) for WASP 76b. The different intensities of red for $B_{x}$ indicate the magnitudes at each time as indicated in the legend, until t/$t_{*}$ $\sim$ 7000. The blue dashed line corresponds to the solution found in case of winding and Ohmic only, without the Hall and ambipolar terms (in which case, by construction, there is no $y$ component in the induction equation, i.e. $B_y=B_y^{\rm in}=B_d$ and $J_x=0$). Note that at the outer boundary there is a sharp discontinuity in both magnetic field components, corresponding to a tangential current sheet (here cut out for visualization purpose). This is due to the (conservative) boundary conditions at the top; it only affects the very last points and should be regarded as a numerical artifact rather than a physical current sheet.}
\label{fig:wasp76b3}
\end{figure}
\begin{figure}
\centering
\includegraphics[width=.9\linewidth]{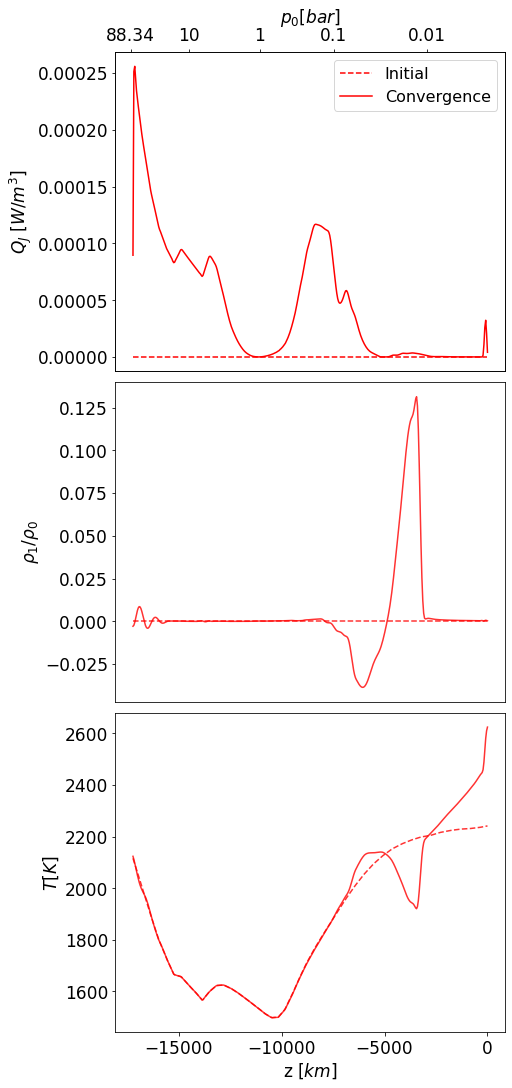}\\
\caption{The vertical profiles of $ Q_{j}$(z) (top), $\rho_{1}/\rho_{0}$(z)(centre) and  $T$(z) (bottom). The dashed line corresponds to the background profile at the beginning of the simulations and the continuous line at the convergence at t/$t_{*}$ $\sim$ 7000. }
\label{fig:3.1second}
\end{figure}

We begin with a detailed analysis of the simulation WASP 76b, which we will take as the reference case. In Fig.~\ref{fig:wasp76b3} we show, from top to bottom, the 1D evolution (marked by increasing color shades) of the vertical profiles $B_{x}$, $B_{y}$, $J_{y}$, and $J_{x}$. The solution converges at a time $t/t_{*} \sim 7\times10^3$, where $t_{*}$ is a proxy to the sound  crossing time over a scale height, approximating the latter as $D/10$ in our simulations (see \citealt{soriano23} for detailed definitions in the isothermal case).

Let us start by examining $B_{x}(z)$, which is initially zero. It quickly increases until convergence, when, at first order, the advection and the Ohmic terms balance each other. At such equilibrium solution, $B_{x}(z)$ has a maximum value $B_x\sim 80$ G at $p\sim 1$~bar, and a minimum value of $B_x\sim - 180$ G at $p\sim0.04$ bar, corresponding to the shear regions, i.e., where the wind profile presents the strongest vertical variations (see Fig. \ref{fig:wind,pressure}). There is a change of sign in the magnetic field at $p\sim0.1$ bar, due to the combination of the change of sign the shear term and the relative flattening of $\sigma(z)$, both at $p\gtrsim 0.3$ bar. 
Note that the locally induced field is $\sim$1-2 orders of magnitude larger than the background field, in line with the non-linear regime expectations in the presence of high temperature \citep{dietrich22}. The winding effect is supported by the meridional current profile, $J_{y}$. As shown in the third panel, it has a minimum of around $p\sim 0.1$ bar and a maximum at $p\sim0.01$ bar. The change of sign and steep profile of $J_y(z)$ is related to the meridional loop which support the meridional induced field.  A comparison between the equilibrium solution with (red darkest line) or without (blue dashed line) the Hall and ambipolar terms shows that the azimuthal component of the magnetic field is mostly set by the winding-Ohmic balance alone, with only slight corrections by the non-linear terms (a few $\%$).

 However, an interesting result is that, at equilibrium, the $B_{y}$ profile strongly deviates from its initial, constant value $B_y^{\rm in}=B_d=3$ G, as seen in Fig. \ref{fig:wasp76b3} (second panel) if the non-linear terms are activated (otherwise, it cannot change by construction).
 Note that, due to the outer BCs, $B_{y}$ in the upper part of the domain is forced to its initial value. However, the overall trend is towards a constant $B_{y}$ approaching zero in the outer layers just beneath the outer boundary.
Correspondingly to the induced $B_y$, there is an azimuthal induced current component, $J_{x}$, shown in the bottom panel of Fig. \ref{fig:wasp76b3}, which is significantly smaller compared to $J_{y}$ throughout the entire domain. For $p \gtrsim 0.5$ bar, its contribution is practically negligible in comparison to $J_{y}$. However, it shows a positive peak at $p \sim 0.08$ bar. This increase is caused by the strong local variation of $B_y$.
In order to understand this behaviour, it is important to notice that the induced Lorentz force triggers non-zero meridional ($v_y$) and vertical ($v_z$) velocities (on the azimuthal one there is no feedback since we force it to remain equal to its input value), via the momentum equation. Both components, which can reach peaks of $\lesssim 10$ m/s, are much smaller than the $v_w\sim $ km/s values, which is consistent with the small but non-zero vertical velocities found in GCMs (e.g., \citealt{rauscher10}). While $v_z$ appears due to the winding-dominating $J_yB_x$ term, a non-zero $v_y$ appears solely as an indirect consequence of the Hall drift, i.e. if $J_x\ne 0$, through the $B_zJ_x$ term in the momentum equation. Such velocities play an important role, since they enter as further advective terms in the induction equation, as we will see below.

Therefore, even though the direct Hall contribution to the induction equation (the term $\propto\partial_z (B_zJ_y)$) is never dominant, its presence triggers a meridional field and a Lorentz force which has a feedback on the fluid and renders the system non-linear.

This is very important, since previous studies \citep{perna10a} were considering the relevance of Hall and ambipolar terms using only the background field to estimate the magnitudes, under the implicit assumption of a perturbative regime (induced field much smaller than the background one). In our simulations, where the winding grows the field to a strongly non-linear regime, the induced field in the meridional direction, $B_y$, is comparable to the background field, because the winding-induced current component $J_y$, which enters as $J_yB_z$ in the $y$-component of the induction equation, is locally very strong.

\begin{figure*}
\centering
\includegraphics[width=.32\linewidth]{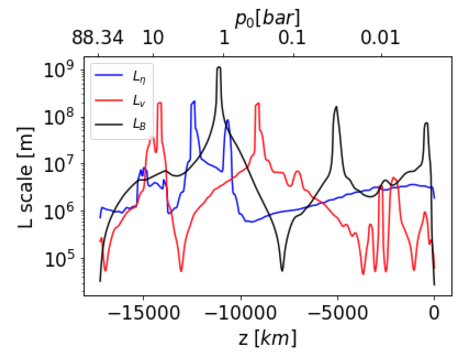}
\includegraphics[width=.32\linewidth]{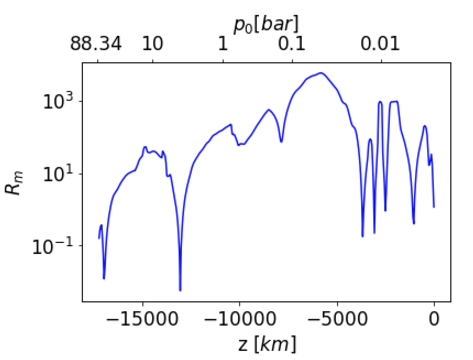}
\hfill
\includegraphics[width=.32\linewidth]{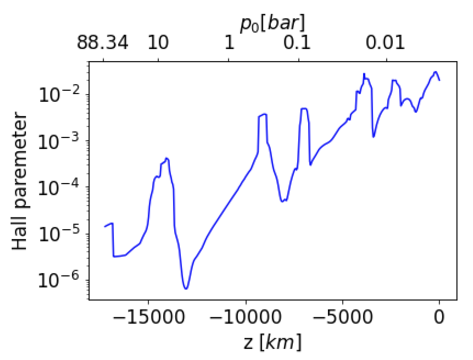}
\caption{{\em Left:}
Local lengthscales calculated with the local vertical gradients of magnetic diffusivity (blue), velocity (red) and induced magnetic field (black) for WASP 76b, at convergence. {\em Center:} magnetic Reynolds number, calculated using $L$ as the local minimum of the three lengthscales shown on the left. {\em Right:} Hall parameter (see text for definition employed) in the studied domain for the default simulation WASP 76b at convergence.}
\label{ans:RmHP}
\end{figure*}

We further calculated the Ohmic heating rate $Q_{j}$, associated with the induced currents. In Fig. \ref{fig:3.1second} (top) it can be observed that most of the Ohmic dissipation occurs for pressures higher than 0.1 bar. Moreover, the highest values are located at pressures higher than 5 bar, which is relevant as it indicates that $Q_{j}$ has the highest contribution in the deepest regions. As a result, more dissipated energy will be available near the RCB; if this energy penetrates into the convective region, it could be redistributed throughout the planet and inflate it more efficiently. The vertically-integrated 
Ohmic dissipation (per unit surface) is  $\int_D Q_{j}~dz=0.0008$~MW/m$^{2}$, which is $\epsilon \simeq 0.016\%$ of the stellar irradiation $F_{\rm irr}$. 

We also examined the influence of magnetic terms on the density and temperature within the region of interest, looking at the deviations from the background profile. The center panel in Fig. \ref{fig:3.1second} illustrates the ratio $\rho_{1}/\rho_{0}$, i.e. the perturbed density over the background density. It shows that magnetic terms induce slight, but non-negligible modifications, to the background density profile. For $p\gtrsim 0.5$ bar, these variations are at maximum $\sim 2\%$, becoming more pronounced at lower pressures, where the density is inherently lower. In these regions, changes can reach up to $\sim 5–12\%$. On the other hand the initial profile of \(T\), shown in Fig.~\ref{fig:3.1second} (bottom), maintains the same shape for \(p \gtrsim 0.1 \, \text{bar}\). However, due to the low density in the upper layers and the ease of energetically modifying these regions, significant variations between the initial and evolved profiles can be observed for $p \lesssim 0.1$ bar. These variations lead to temperature changes locally exceeding \(200 \, \text{K}\).
 Looking at the contributions to the energy equation, the forcing and cooling term balance each other everywhere, except in the outer regions where the buoyancy-related term ($\rho g v_z$) becomes significant and causes thermodynamic deviations from the background profiles, at equilibrium. The Ohmic rate is never dominant, and therefore its heating is not reflected. Due to the sensitivity of the results on the forcing, the cooling term, and the profile of $v_z$ (i.e., on the damping term in the $S_z$ evolution equation in the outer region, see above), the $T$ profile should be taken with much caution.

Finally, we have calculated the magnetic Reynolds number, $R_{m}$, and Hall parameter to understand the relative contributions of the advective and Hall terms compared to the Ohmic term. There is however a subtle point which is important to discuss. Since we are looking for exactly stationary solutions, $\partial_t \mathbf{B}=0$, which is a balance largely dominated by the advective and Ohmic terms, the formal ratio of the contributions to the induction equation (see e.g. eq. 4 of \cite{dietrich22}:
\begin{equation}
\label{eq:rm_curls}
R_m=\frac{|\nabla\times(\textbf{v}\times\textbf{B})|}{|\nabla\times(\textbf{J}/\sigma)|}
\end{equation}
by definition is $\sim 1$ at equilibrium, regardless of the velocity and of ratio between the background and induced magnetic fields. This may appear puzzling, but it is reconciled with the expected non-unity values of $R_m$ by noticing that the usually definition simplifies the ingredients to:
\begin{equation}
\label{eq:rm_simple}
R_m \simeq \frac{v B_{bkg}}{L_{v\times B}}\frac{L_{\eta}L_{B}}{\eta B_{ind}} \simeq \frac{v L}{\eta}\, ,    
\end{equation}
where $\eta=1/(\sigma \mu_0)$ is the magnetic diffusivity. This definition assumes that $B_{ind}\simeq B_{bkg}$, which is not the case when $B_{ind}\gg B_{bkg}$ (reminding that $B_{ind}\perp v$ so $v\times B\sim v\times B_{bkg}$, and $J\sim B_{ind}/L_{B}$), so that $v\times B\sim v B_{bkg}\ll v B_{ind}$. This means that the standard approximation, eq. (\ref{eq:rm_simple}), results in $R_m$ being an order-of-magnitude estimate for the ratio between the induced and background fields.
However, even neglecting this subtlety (i.e., assuming $B_{ind}=B_{bkg}$ in the definition of $R_m$), the calculation of the Reynolds number presents the ambiguity of the definition of the lengthscale $L$: the velocity, the magnetic diffusivity, and the magnetic field vary non-trivially, and the local associated length scales are in general comparable and oscillating, as seen in the left panel of Fig. \ref{ans:RmHP}. None of them is negligible in all the domain. In any case, in order to have a useful quantification of $R_m$ in the standard way, the middle panel of Fig. \ref{ans:RmHP} shows $R_m$ using $L(z)=$min($L_{v}$,$L_{\eta}$,$L_{B}$), as done by \cite{dietrich22}. We observe that $R_m$ remains greater than 1 in practically the entire domain, a characteristic of the advection-dominated (non-linear) regime studied in this paper.

We can also evaluate the Hall parameter (defined as $\omega_e \tau_e$, where $\omega_{e}$ is the electron gyrofrequency and $\tau_e$ the electron collision time), by showing directly the ratio between the Hall and Ohmic contributions,
\begin{equation}
 \omega_e \tau_e = \frac{\left|\nabla\times\left(\frac{\textbf{J}\times \textbf{B}}{en_{e}}\right)\right|}{|\nabla\times(\textbf{J}/\sigma)|}\,.   
\end{equation}
In the right panel of Fig. \ref{ans:RmHP}, we can see that the Hall parameter remains between $10^{-6}$ and $10^{-2}$, which is indicative of the dominance of the Ohmic term over the Hall term. However, note that as the pressure decreases, the Hall parameter increases, indicating the growing relevance of the Hall effect in the outermost layers. Again, it is important to note that the analytical estimate of the Hall parameter is also affected by specific choices of length scales and velocities. In any case, the Hall parameter (or magnetization) remains much smaller than one, which is expected, see e.g. the small magnetization parameters in the M1 region in \cite{koskinen14}.

\subsection{Comparison between modeled Hot Jupiters}
\label{subsec:difplanets}
\begin{figure}
\centering
\begin{adjustbox}{margin=0pt 0pt 5pt 0pt}
\includegraphics[width=.9\linewidth, trim=0 0 0 0pt, clip]{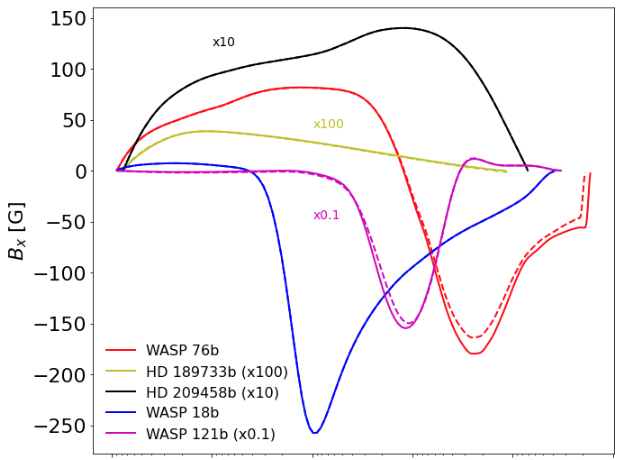}
\end{adjustbox}
\begin{adjustbox}{margin=10pt 0pt 0pt 0pt}
\includegraphics[width=.88\linewidth, trim=0 0 0 0pt, clip]{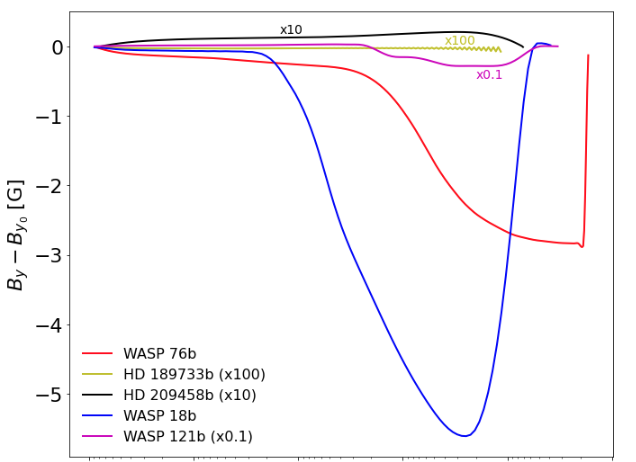}
\end{adjustbox}
\begin{adjustbox}{margin=7pt 0pt 0pt 0pt}
\includegraphics[width=.925\linewidth, trim=0 0 0 0pt, clip]{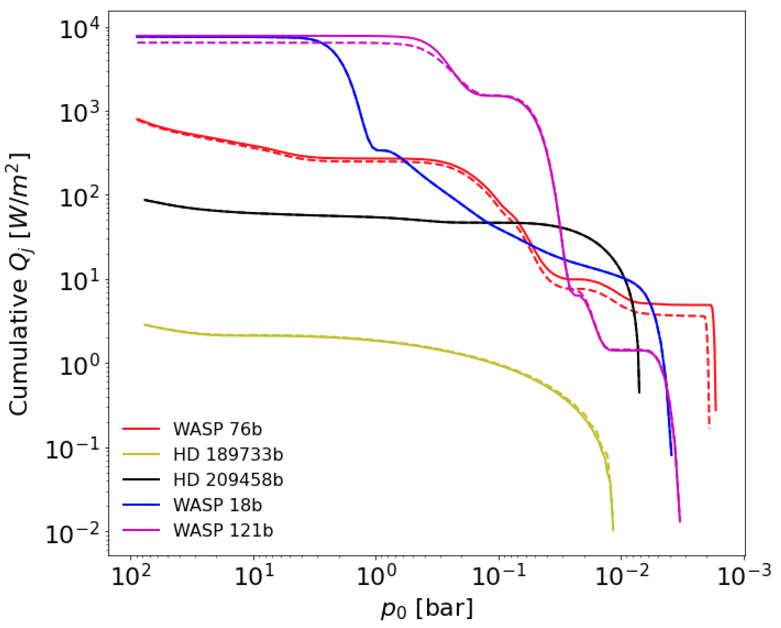}
\end{adjustbox}
\label{fig:differentplanets}
\caption{Comparison of the vertical profiles (from top to bottom) $B_{x}(z)$, $B_y(z)$, and cumulative Ohmic heating, $\int_z^0 Q_j(z')~dz'$, for the different models. The dashed lines correspond to the purely winding+Ohmic cases, with no Hall and ambipolar terms included.}
\label{fig:differentplanets}
\end{figure}

Let's now compare the solutions for the different planets, shown in Fig. \ref{fig:differentplanets} and Table~\ref{table2}. In general, there are two competing temperature-dependent effects for sustaining the magnetic field. On one hand, we expect that a hotter planet will have stronger temperature gradients and hence faster zonal winds to amplify the magnetic field, and higher values of $\sigma$, thus higher induced fields. On the other hand, there is a larger counteracting magnetic drag expected, which will slow down the wind. In this study, we can only evaluate the first effect, since the second one requires, as already mentioned, a fully consistent MHD treatment in GCMs.

A close inspection of the simulation results for
the five planets shows that the one presenting the highest induced field is WASP 121b, with a maximum absolute value of the main field component $|B_x|_{\rm max}\sim 1550$~G. Although we might have expected WASP 18b to have a higher magnetic field compared to the previous one based on its $T_{eq}$ and the highest initial field, we rather find that it is lower than
WASP 121b, with a maximum of $|B_x|_{\rm max}\sim 260$~G. This is
because the magnetic drag present in the winds used as input from the GCM is based on a field strength of $B_{d}$ = 20~G, higher than the $B_{d}$ = 3~G of the other planets, which substantially damps any wind at pressures p $\lesssim$ 1 bar (Fig.\ref{fig:wind,pressure}). Following the two aforementioned planets, we have WASP 76b, with a maximum $|B_x|_{\rm max}\sim 180$~G. The colder HJs HD
209458b and HD 189733b show much lower induced fields: $|B_x|_{\rm max}\sim 14$~G
and only $\sim$ 0.4 G, respectively. Note that, even in these cases, the locally induced field is of the same order of magnitude of the background field (at least at the sub-stellar point), making the linear regime assumption $|B_x| \ll B_{d}$ inadequate. 

Fig. \ref{fig:differentplanets} also shows a comparison of the results including only the Ohmic and winding terms, excluding the Hall and ambipolar contributions (dashed lines, to be compared with the full solution with solid lines).
Minor differences in $B_x$ can be seen for the hottest planets, WASP 76b and WASP 121b, where there is a slight correction introduced by the non-linear terms (Hall and ambipolar), as previously discussed for WASP 76b. These differences are noticeable in the upper part of the domain due to the increase of the contribution of the ambipolar and Hall terms in the region. However, for the cooler planets, the contribution of the Hall and ambipolar terms is significantly smaller, as will be further discussed in the next section, and they produce virtually no differences in the $B_x$ profile.

On the other hand, the Hall and ambipolar terms also affect the evolution of $B_y$.
This component evolves over time, deviating from its initial value as it reaches equilibrium, particularly in cases where the Lorentz force varies steeply with altitude, as seen in the central panel of Fig. \ref{fig:differentplanets}. Typically, $B_y$ reaches peak values of the order of $\sim$ G, as shown in detail for WASP 76b. For the other two hot planets, WASP 121b and WASP 18b, $B_y$ reaches similar or even slightly higher values compared to WASP 76b. In contrast, for the colder planets HD 209458b and HD 189733b, $B_y$ remains nearly unchanged from its initial value of 3 G. This is due to the much weaker Hall and ambipolar effects, resulting in minimal evolution of the magnetic field. In these cases, $B_y$ remains practically constant throughout the entire domain.

If we now examine the corresponding accumulated dissipated energy, $\int_z^0 Q_j(z')~dz'$ (bottom panel of Fig. \ref{fig:differentplanets}), we observe that the planet with the highest dissipation are WASP 18b, WASP 121b and WASP 76b. These hot planets exhibit Ohmic dissipation rates that are approximately 1 to 3 orders of magnitude higher than those of the colder planets HD 209458b and HD 189733b, respectively.
When comparing the cumulative $Q_j$ profiles for each planet, with and without the inclusion of the Hall and ambipolar terms, we find that the differences between the two cases are minimal. However for those simulations with higher contribution of the Hall and ambipolar terms, WASP 121b and WASP 76b, the cumulative $Q_{j}$ is slighlty higher compared to the models with just advection and Ohmic terms.

In all cases, most of the energy is dissipated around the shear layer, so that the cumulative energy saturates at a pressure $\sim 0.1-$few bars, depending on the planet. If we analyse the local heating efficiency for the different planets (eq.~\ref{eq:efficiency}), we can see that the highest value is reached for WASP 121b, $\epsilon=0.1\%$, compared to the deposited energy from the star, while the lowest is HD 189733b, $\epsilon=0.0009\%$. Compared with the statistical study by \cite{thorngren18} which infers a maximum heating efficiency for $T_{\rm eq}\sim 1600$ K, after which the efficiency quickly drops, here we see a more monotonic trend $\epsilon(T_{\rm eq})$ for two main reasons: (i) our local simulations only allow an estimate of the local efficiency at the substellar point which is substantially hotter than the average, so that the local estimate is highly overestimating the global efficiency; (ii) since the induction is non-linear and $B_x\gg B_d$ especially at high $T_{\rm eq}$, the magnetic drag GCM is probably largely underestimating the real effects that slow down the winds. Moreover, \cite{thorngren18} and similar works consider only the heat deposited below the radiative-convective boundary, therefore our results, confined to the uppermost, radiative layers, are not directly comparable to theirs (see e.g. \citealt{batygin11,wu13} for the expected radial profiles of Ohmic heating rate across all planetary layers).

\subsection{Contributions to the electric components}
\label{subsection:hall}
In order to properly quantify the relative effect of the advective, Ohmic, Hall, and ambipolar terms in our simulations, in this section we present an analysis of their contribution to the electric field components which govern the induction equation. 

The analysis of the representative cases WASP 76b and HD 209458b is presented in Fig.~\ref{fig:magneticterms} (top and bottom panels, respectively), showing the different contributions to the electric field components $E_x$ (left) and $E_y$ (right). These contributions depend on the velocity, magnetic and current fields and on the prefactors, as shown in Eq.~(\ref{eq:induction}), which are determined by the local density and temperature. Note that $E_{x}$ and $E_{y}$ are flat in all the domain (except for the very last points, a boundary artifact), indicative of the stationary state. In general, several dips appear in the magnitude of most contributions, corresponding to changes of sign in the dominant components of each term. Both the Hall and ambipolar contributions increase in altitude, especially at the high pressures of the domain, due to their density dependence.

For WASP 76b, the analysis of the $E_y$ component (top right), which is primarily linked to the induced magnetic field $B_x$, shows that the dominant contributions arise from the Ohmic and advective terms (i.e., magnetic winding). As expected, the Hall and ambipolar contributions are several orders of magnitude smaller throughout the entire domain. Their contributions increase at lower pressures due to the decreasing density; however, they remain significantly smaller than the dominant terms, with the exception of the ambipolar term $\propto B_{x}B_{x}J_{y}$, which can locally reach values comparable to the advective term $B_{z}v_{x}$ for $p < 0.01$~bar. However, at these layers, the vertical advective term $B_xv_z$ is the one balancing out the Ohmic dissipation. For the $E_x$ component (top left), the Ohmic term is apparently flat above $p \gtrsim 0.04$ bar, and its small derivative is balanced by the ($B_{y}v_{z}$) and Hall ($J_yB_z/(en_e)$) components. At shallower layers, the derivative of the Ohmic and the Hall terms dictate the equilibrium, because the absolute value of the $v_zB_y$ contribution is larger, but constant.
\begin{figure*}
\centering
\begin{minipage}[t]{0.48\linewidth}
    \centering
    \includegraphics[width=\linewidth]{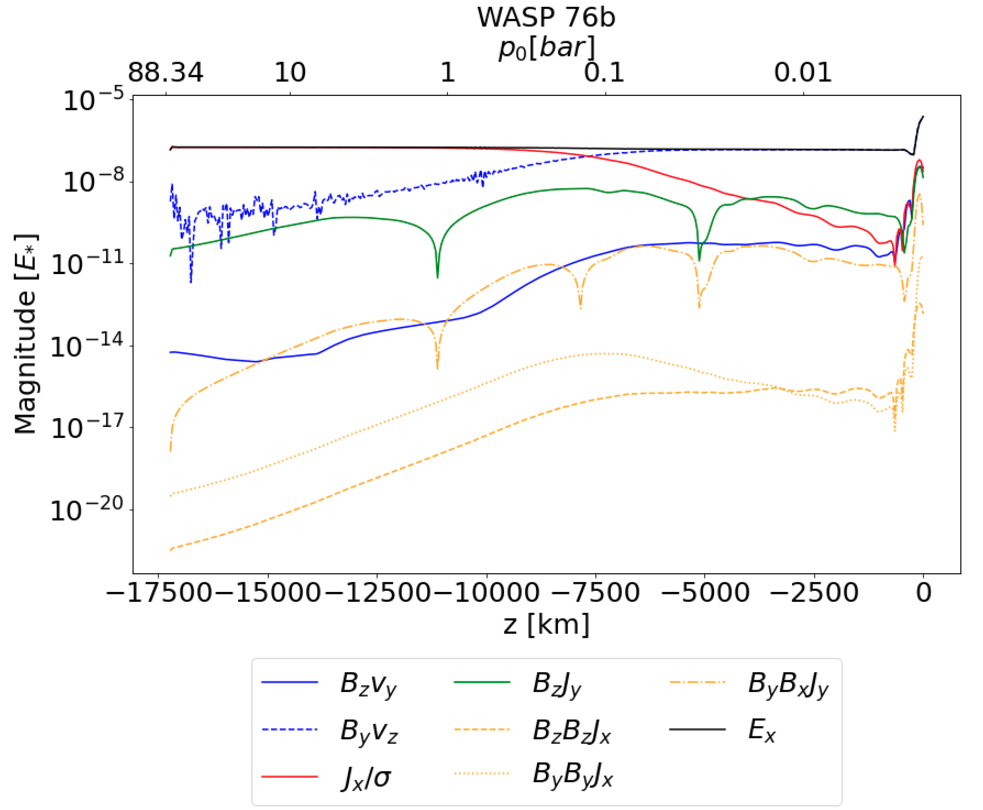}
\end{minipage}
\hfill
\begin{minipage}[t]{0.48\linewidth}
    \centering
    \includegraphics[width=\linewidth]{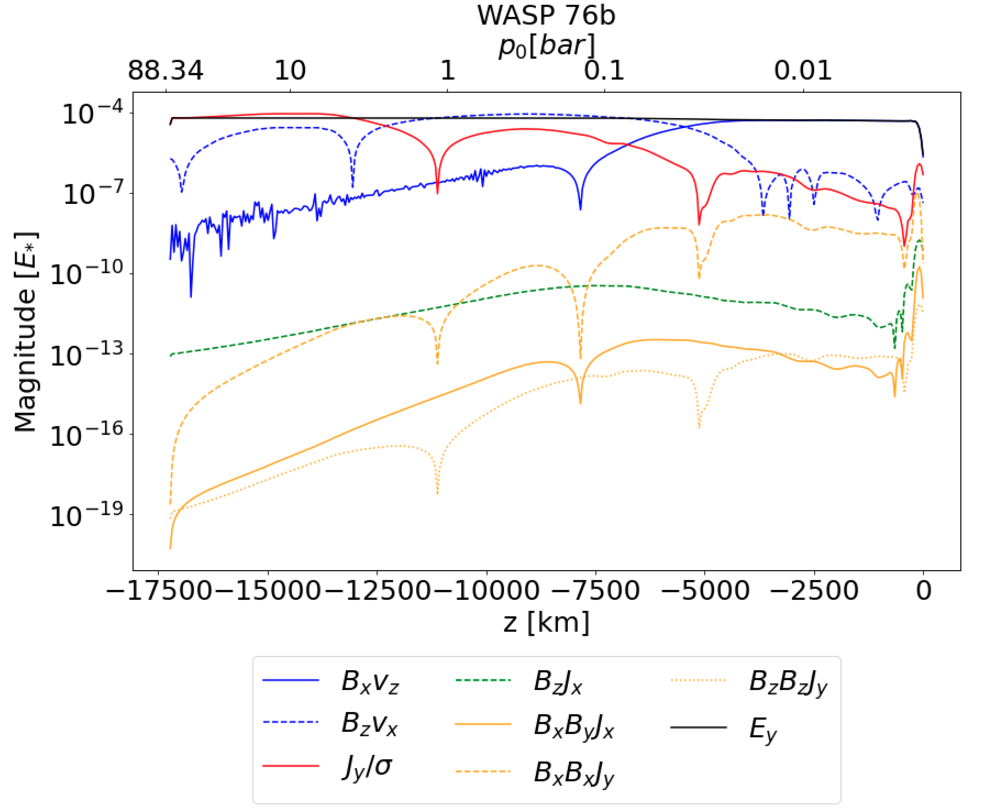}
\end{minipage}

\vspace{1em} 

\begin{minipage}[t]{0.48\linewidth}
    \centering
    \includegraphics[width=\linewidth]{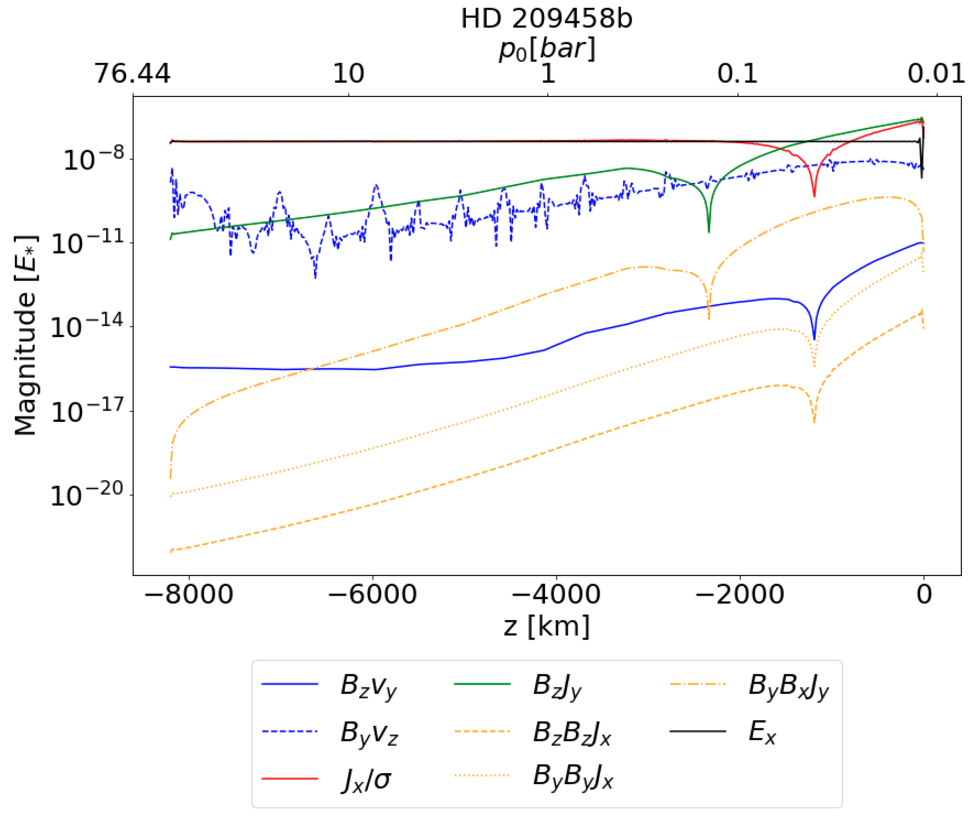}
\end{minipage}
\hfill
\begin{minipage}[t]{0.48\linewidth}
    \centering
    \includegraphics[width=\linewidth]{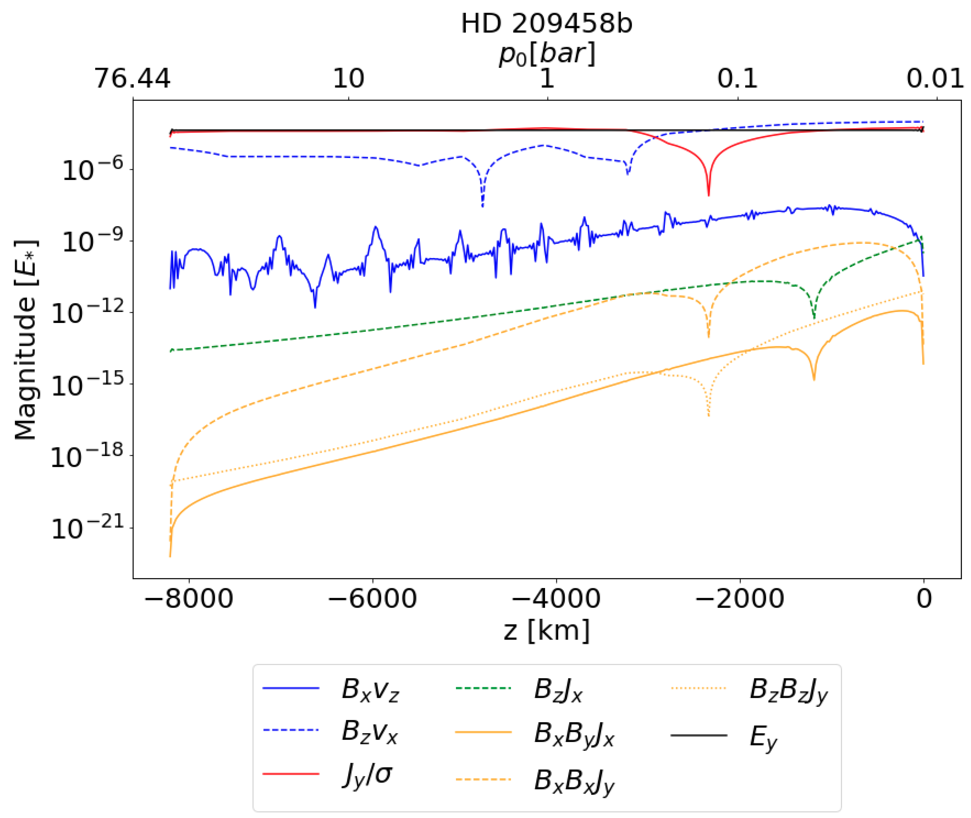}
\end{minipage}
\caption{Evaluation of the different contributions to the electric field components: $E_{x}$ (left) and $E_{y}$ (right) for WASP 76b (top) and HD 209458b (bottom) across the entire domain. Advective terms are shown in blue, Ohmic in red, Hall in green, ambipolar in yellow, and the total electric field ($E_{x}$ and $E_{y}$) in black. The continuous, dashed, and dash-dot lines represent the non-zero different components of each contribution, as in the legend (which omits for brevity the Hall and ambipolar pre-factors).}
\label{fig:magneticterms}
\end{figure*}

For HD 209458b, $E_{x}$ represents a balance between the Ohmic, the advective $v_{y}B_{z}$ and the Hall terms, with the advective $B_{z}v_{y}$ term always playing a minor role. Since the largest, Ohmic term is practically constant, the advective and Hall terms also play an important role in the evolution of the $B_{y}$ component. However, the order of magnitude of the terms is slightly smaller compared to WASP 76b, thus almost no significant variations of $B_{y}$ are expected, as seen in the previous section for the cooler planets. Regarding the component $E_{y}$, we observe that, for all the domain, the main contributions come from the advective term ($B_{z}v_{x}$) and the Ohmic term. Although the Hall and ambipolar terms increase as pressure decreases, they remain several orders of magnitude smaller than the Ohmic and advection terms, and smaller by one order of magnitude compared to the same contributions for WASP 76b. The analysis of HD 209458b is representative of the cooler HJ models, and shows the general behaviour of the different components that play a role in the magnetic field evolution equation.

\subsection{Ion-neutral relative velocity and its detectability}

Our results on the ambipolar term can be connected to the recent work by \cite{Savel_2024}. They proposed a novel method to constrain the magnetic field in hot gas giants by comparing the velocities of heavy ions and neutral gas using high-resolution spectroscopy. Ideally, if one measured the ambipolar velocity, eq.~(\ref{eq:v_amb}), in the photospheric region, the magnetic field of the planet could be constrained.

In Fig.~\ref{fig:Savel}, we plot the profiles of the relative velocity, inferred from our simulations. Note that our domain is much deeper than the photosphere, $p\lesssim 10^{-4}$ bar, which is where one could observationally test predictions. We observe a common trend for all planets. First, for the highest pressures achieved in our simulations, the relative velocity difference is negligible, but grows by several orders of magnitude outward, due to the decrease of both $n_{p}$ and $\nu_{in}$. In the bulk of our domain ($p\gtrsim 0.01$ bar), the values of the relative velocities are still very small, but our results, at these depths, are roughly in line with the calculations by \cite{Savel_2024} (see their Fig.~1, especially in the 150~G magnetic field line, a value which is the same order of magnitude of the induced magnetic field in our simulations).

However, there are important differences, compared to the smooth, monotonically growing outward profiles by \cite{Savel_2024}. First, the noticeable drops in the intensity of the velocity difference, at specific pressure values, correspond to the changes of sign in the ambipolar term. 
Secondly, and more importantly, we observe either a decrease in the relative velocity for WASP 121b, WASP 76b, WASP 18b and even HD 209458Bb, or a reduction in its rate of increase (i.e., a flattening of the slope) for HD 189733b, at the lowest pressure levels. This is mainly due to the increasing relative importance of the ambipolar diffusion itself compared to the winding term, as seen in Fig.~\ref{fig:magneticterms}, which increases exponentially with pressure and their effect is visible in WASP 121b and WASP 76b. The ambipolar term tends to minimize the Lorentz forces and limits the growth of the currents perpendicular to the magnetic field, which are mostly induced by winding ($J_y$ and $B_x$). However, our enforcement of no induced field at the boundaries (see App.~\ref{appendixA}) could also have an effect, so that one should ideally extend outwards the domain to have a better assessment (which is numerically unfeasible due to the too large top/bottom density contrast and low values of density and pressure, see App. A of \citealt{soriano23}). Moreover, we are employing values of background magnetic field much lower than those used in \cite{Savel_2024}: higher values of $B_d$ could enhance the relative velocities.

\begin{figure}
\centering
\includegraphics[width=0.9\linewidth]{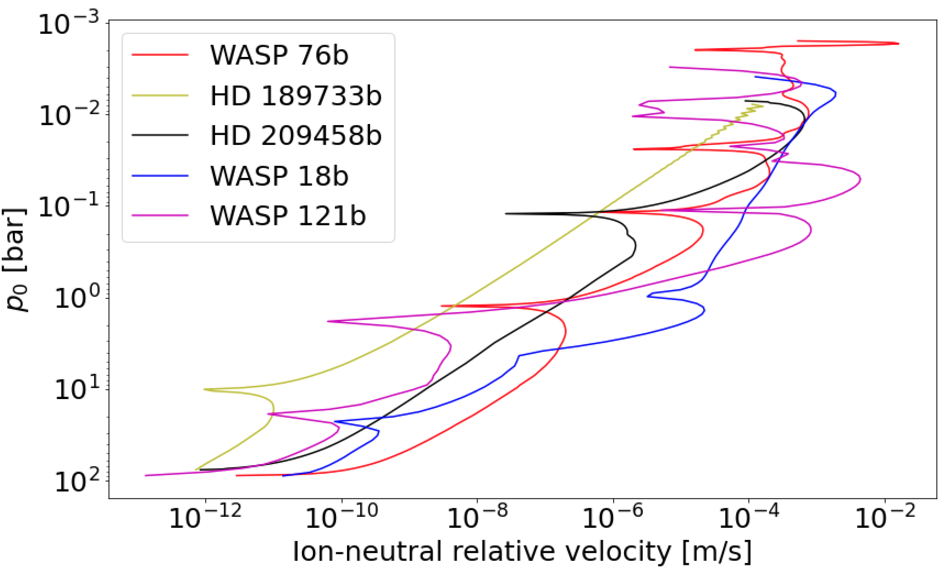}\\
\caption{Relative velocity between ions and neutrals inferred from the solutions of the different models}.
\label{fig:Savel}
\end{figure}

Although, for the reasons just discussed, we cannot give a clear quantitive confirmation of the potential spectroscopical detectability of the ion-neutral relative velocity in the photosphere (needed to be $\gtrsim$ km/s, \citealt{Savel_2024}), we confirm that the ambipolar diffusion is not negligible in the uppermost layers, and it is intimately related to the other terms and the configuration of the background and induced magnetic fields.

\subsection{Sensitivity on the maximum depth considered}
\label{subsec:depth}

The depth at which most of the energy is deposited is relevant for the inflation of the planet. The deeper the energy is dissipated, the less energy is needed to inflate the planet \citep{guillot10}, with dissipation below the radiative-convective boundary being optimal. Inflation efficiency reaches its maximum if the energy is deposited in the convective region, allowing the energy to be redistributed throughout the entire planet \citep{ginzburg16}. Since in our approach, by construction, we confine the induction to the numerical domain only, in the following we analyse the effect of extending the domain to deeper regions for the reference case of WASP 76b (see \S \ref{sec:extension}). Three different extensions have been studied, with $p_{\rm max}$= 300, 500 and 1000  \rm{bar} respectively; see Table~\ref{table2}. Also, note that the top pressure has been kept fixed at  $p_{\rm top}$=0.005~\rm{bar}.

\begin{figure}
\centering
\begin{adjustbox}{margin=2pt 0pt 0pt 0pt}
\includegraphics[width=.905\linewidth, trim=0 0 0 0, clip]{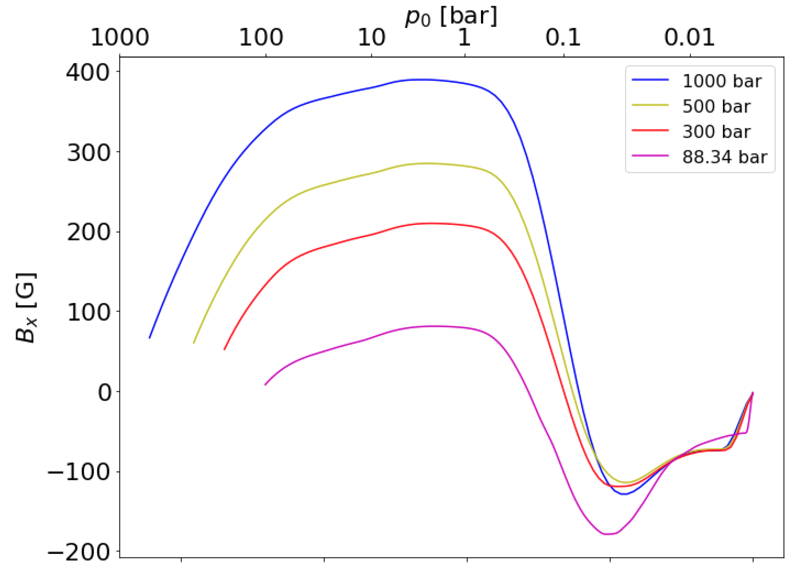} 
\end{adjustbox}
\includegraphics[width=.925\linewidth, trim=0 0 0 0, clip]{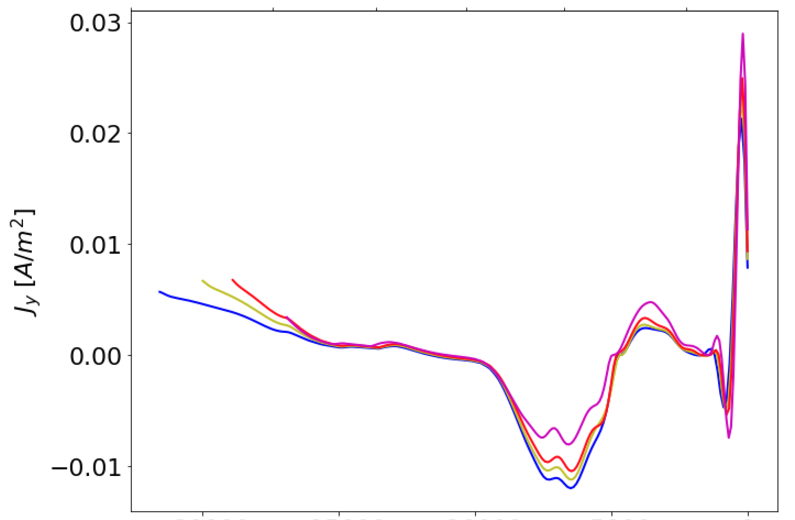}\\
\begin{adjustbox}{margin=2pt 0pt 0pt 0pt}
\includegraphics[width=.89\linewidth, trim=0 0 0 0pt, clip]{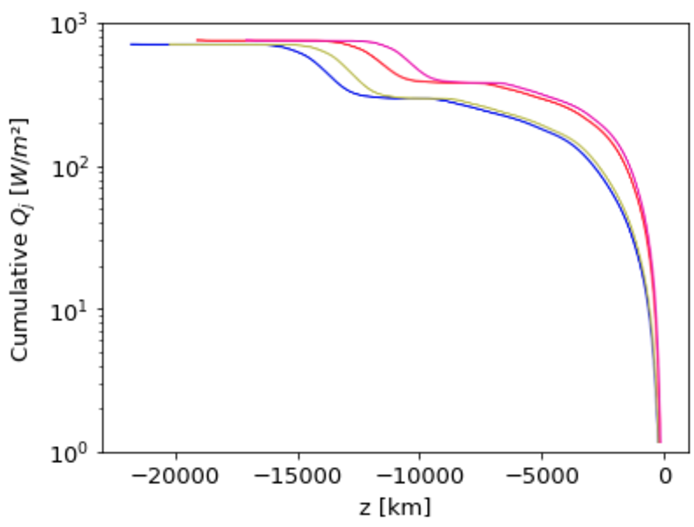}
\end{adjustbox}
\caption{The vertical profiles for  $B_{x}$(z) (top), $J_{y}$(z) (center) and cumulative Ohmic heating $\int^0_z Q_{j}(z')dz'$ (bottom) at convergence, corresponding to $t/t_{*}\sim 10000$ for four simulations with different maximum pressures: 1000 bar (blue), 500 bar (yellow), 300 bar (red) and 88.34 bar (pink) for WASP 76b. The minimum pressure is the same at the top of the four simulations, 0.005 bar. In these simulations, we only considered winding and Ohmic dissipation.}
\label{fig:DifferentP}
\end{figure}

For simplicity, we focus on the dominant induced components $B_x$ and $J_y$, and consider cases without the Hall and ambipolar diffusion, which, in any case, are completely negligible in the extended deep regions. When analysing differences in the magnitudes of the relevant simulated variables for various values of the maximum pressure, we observe the following trends. First, as the maximum pressure increases, the generated magnetic field strength also increases (see Fig. \ref{fig:DifferentP} top). This is essentially due to the chosen boundary conditions, which tie $B_x$ at both extremes, allowing to reach higher peaks if the domain is more extended. The highest field strength is observed close to the shear layer, e.g. at $p\simeq 1$ bar for the most extended simulation ($p_{\rm max}=1000$ bar), corresponding to approximately 390~\rm{G}. Notably, the vertical variability of the magnetic field is similar across the four simulations, with the maximum values for $B_{x}$ consistently around 1~\rm{bar}, as explained in \S\ref{sec:resultsintro}.

Examining the amount of currents generated, $J_{y}$ (see middle panel of Fig. \ref{fig:DifferentP}), we note that, despite the shifts in pressure across different simulations, the overall behaviour remains consistent with the patterns described in \S\ref{sec:resultsintro}. The currents exhibit positive values below 1 bar, negative values between $\sim 0.05$ and $\sim 1$ bar, and positive values below $\sim 0.01$ bar. Moreover, the deeper is the lowest boundary of the simulation, the highest is the absolute value of the minimum of $J_{y}$, since there is a steeper increase of magnetic field in the shear region.

Finally, we study the cumulative Ohmic dissipation $Q_{j}$ for the different cases (see Fig. \ref{fig:DifferentP}, bottom). We note that the main contribution to dissipation appears to be largely concentrated around a few bars. As a matter of fact, in the outermost region, as the pressure increases, the amount of accumulated Ohmic rate rises rapidly until it reaches a constant value, which is similar for all simulations.
On the other hand, after this constant value is achieved, there is another significant increase in $\int_z^0 Q_{j}(z')dz'$ for all cases until the four simulations gradually converge to a nearly identical total deposited energy of $7.8-8.1\times10^{-4}$~ MW/m$^{2}$. Even for the most extended case, the maximum cumulative energy is reached at pressures around $\sim 50$ bars. Therefore, extending the domain does not contribute to a higher cumulative energy.

This convergence with the domain size might be an effect of fixing $B_x=0$ at $p_{\rm bot}$. When this condition is relaxed, and a global solution is retrieved, the induced field and currents, and the associated Ohmic dissipation, partially spread to deeper layers, as long as the conductivity is non-zero \citep{batygin10,batygin11,wu13,knierim22}. In this sense, in our study, we cannot quantifying the global Ohmic dissipation, also because, as said above, we have a local setup at the sub-stellar point, and not a fully-consistent GCM with the magnetic feedback on the fluid.

\subsection{Sensitivity on the internal magnetic field}
\label{subsec:magneticeffect}

In this section, we explore the effect that the strength of the radial background magnetic field component $B_z$ has on the induced field, since the growth due to winding is proportional to it. Variations in the magnitude of the $B_{z}$ component can be either due to a different overall strength of the magnetic field of the planet, to a different alignment, or different multipolar contributions

It is important to highlight that these runs have been performed using the input from GCM models of WASP 76b but without any magnetic drag, and therefore with higher wind speeds. We prefer this choice since inputs from GCM models with drag would implicitly fix a background magnetic field. As a consequence, the induced fields are likely overestimated; however, we are interested in relative comparisons. Furthermore, for this test, we include only the dominant winding and Ohmic terms (therefore, the value of $B_y^{\rm in}$ is irrelevant since it doesn't enter in our setup, i.e. the plane-parallel advective+Ohmic induction equation).

\begin{figure}
\centering
\includegraphics[width=.9\linewidth]{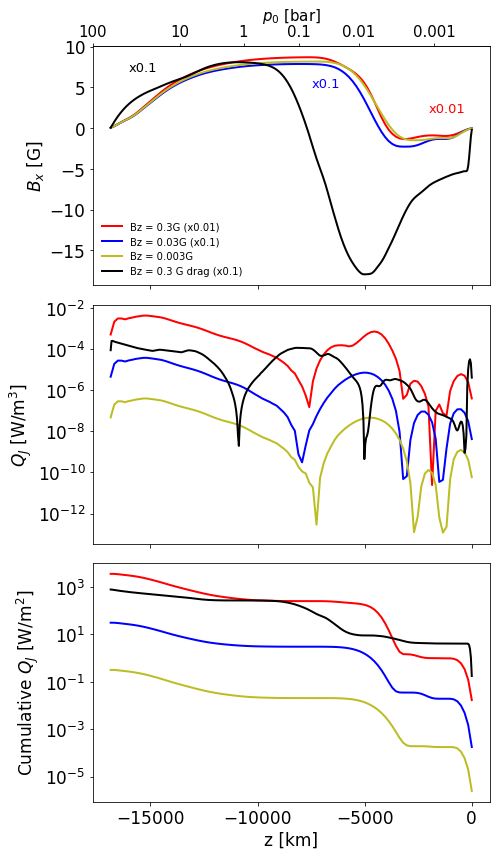}
\\
\caption{Comparison of the  $B_{x}({z})$ (top), the Ohmic dissipation per unit volume $Q_{j}(z)$ and the cumulative one, $\int^0_z Q_{j}(z')~dz'$, for different values $B_z^{\rm in}$=0.3 G (red), 0.03 G (blue) and 0.003 (yellow) for WASP 76b. As a reference, the default case with GCM inputs that consider a drag of $B_d=3$ G is shown in black, for which $B_z^{\rm in}=0.3$ G. Note that some $B_{x}$ profiles have been rescaled as mentioned in the legend, for the sake of clarity. In these simulations we only focus on the dominant winding and Ohmic effects, neglecting the Hall and ambipolar terms.}  
\label{fig:imf}
\end{figure}

In Fig. \ref{fig:imf} we show $B_x(z)$, $Q_{j}(z)$, and the cumulative $Q_{j}(z)$ exploring three different values of the internal field. In the top panel, we see that the induced magnetic field $B_{x}$ roughly scales with $B_{z}$ (as expected in the linear regime), varying between $B_x\simeq8$~G for $B_z=0.003$~G, and 870 G for $B_z=0.3$~G. The maximum values of the induced fields are much higher than the background ones as expected for all cases, typically increasing the initial field by about three orders of magnitude, due to the non-linear regime caused by the high conductivity. 

We clearly note the linearity of the dominant winding effect: the profile perfectly scales with $B_z$, for the three simulations without drag. Consequently, the dissipated heat $Q_j$ and its cumulative value (middle and bottom panels), and the heating efficiency, scale quadratically with $B_{z}^{\rm in}$. In all the cases, the Ohmic dissipation occurs almost entirely at levels $p\gtrsim0.1$ bar, with the bulk of it mainly at $p \gtrsim$1 bar, since the value of the currents is higher in deeper regions.

It is important to highlight that the local efficiency values are likely overestimated for the high magnetic field cases, since the strong induced magnetic field would slow down the wind to speeds lower than the GCM profiles assumed here, which would result in a less pronounced shear layer, and consequently, a lower induced magnetic field. 

In the same Fig.~\ref{fig:imf}, we compare the results with the simulations that include profiles with the drag models (black line), to be compared with the case with the same $B_z^{\rm in}$ (red). For the magnetic field, $B_{x}$ it can be seen that, in the case where magnetic drag is not present, the field $B_{x}$ can increase up to $|B_x|_{\rm max}\sim 870$ G compared to $|B_x|_{\rm max}\sim 180$ G that is reached when the drag is present in the GCM model. This further implies a higher amount of Ohmic dissipation, which accumulated, results in one order of magnitude higher in the case without drag.
Correspondingly, this also implies a more efficient heat deposition for the non-drag case,  of about $\epsilon\sim 0.71\%$, compared with the $0.16\%$ for the case with drag.

\section{Final remarks}\label{sec:conclusions}

In this study, we have performed 1D MHD simulations of HJs atmosphere columns to investigate the impact of the non-ideal MHD effects, besides advection (i.e., winding): Ohmic, Hall, and ambipolar terms. We have employed wind and $p(T)$ profiles obtained from GCM of several specific planets, spanning a range of equilibrium temperatures. The assumed background magnetic field has been parametrized by two constants, the meridional and the radial components. We have solved the MHD equations in their perturbative form, where the velocity and the pressure (i.e., temperature) are forced to keep very close to the prescribed profiles; keeping the full MHD set allows us to include part of the feedback on the fluid (on the meridional and vertical velocities, initially set to zero), and, to a lesser extent, to evaluate the location and magnitude of expected relative deviation from the background profiles. However, the system substantially reduces to the evolution of the induction equation, until it converges to a solution.

The main and common behaviour observed in all the simulations is the generation of a strong ($\sim 10^1$-$10^2$ G), localized azimuthal magnetic field ($B_x$), supported by meridional currents ($J_y$). This is due to the balance between the winding mechanism and the Ohmic term. The induction is most effective close to the shear region, which is the main one contributing to the total Ohmic dissipation along the column, which is typically a fraction $10^{-6}$-$10^{-3}$ of the irradiation. Such heating efficiency is in line with previous studies \citep{batygin10,perna10a}, especially if one considers that, by construction, it includes only the contribution from the thin radiative layers, and not the one in the convective region (which we cannot model with our approach), which is the most important to take into account for HJ inflation studies \citep{komacek17,thorngren18}.

The Hall and ambipolar terms are generally smaller than the dominant winding and Ohmic terms. However, their effect is interesting, especially in the outermost layers of the hottest planets. At second order, the Hall effect modifies the solution, contributing to an additional component of the magnetic field in the meridional direction (i.e., an azimuthal current). Moreover, although the contribution of the Hall term to $E_{y}$ is not dominant, its variation may introduce non-linear effects in the evolution of $B_{x}$ that combined with the ambipolar contribution (which tends to dissipate the currents perpendicular to the magnetic field) modifies the $B_{x}$ profile. Note that this effect is relevant for considerations on the spectroscopic detectability of ion-neutral relative velocity (Savel et al. 2024), and for a proper evaluation of the widely used magnetic drag in GCMs.

We explored variations in several parameters to examine their influence on the winding effect. First, we found that extending the simulation domain to deeper levels, up to 1000~bar, increased the vertical range where the magnetic field can amplify, leading to higher magnetic field strengths. Another critical factor affecting the magnitude of the magnetic field growth due to winding is the planetary internal magnetic field, with a roughly linear relation, as expected from the induction equation $x$ component.

When studying different planets, we quantified how the planetary temperature, i.e. the stellar irradiation, plays a crucial role for multiple reasons. First, the GCM models produce larger velocities for hotter planets; second, the conductivity, dominated by the thermal ionization of alkali metals, is much larger and leads to a much stronger winding effect.

As we conclude, we should note that some words of caution are needed. Our results should not be taken as fully self-consistent, since: (i) they are local simulations of sub-stellar columns, in plane-parallel approximations; (ii) the GCM models used for the input profiles assume an effective magnetic drag term with a pure magnetic dipole aligned to the spin axis, thus implicitly assuming a linear regime for which the induced field is much smaller than the background dipole; (iii) we conservatively confine the induction to the domain considered, imposing no-induced field boundary conditions for practical reasons (allowing a general non-zero induction at the boundaries often leads to an indefinite, non-converging growth of magnetic field); (iv) the magnetic resistivity is assumed to be time-independent, which has minimal impact in the inner domain due to small temperature variations. However, for the outer atmosphere, where temperature differences are significant, slight changes in magnetic diffusivity are expected, as seen in \cite{Menou2012} and \cite{hardy22} and will be considered in future work.

Despite these caveats, our results show how important the non-ideal MHD effects are, and how easily the non-linear regime of the induction can be triggered, where the induced magnetic field is larger than the background one. At the substellar points, even our coldest HJ considered here, $T_{\rm eq}\sim 1200$ K, induce fields locally comparable to the background. The first message is that the  perturbative regime might be appropriate for the low-irradiated end of the HJ sample only, something already stressed e.g. by \cite{batygin13,dietrich22}. Here we found that not only the winding can lead to magnetic fields locally much larger than the background field, but also that the second-order Hall effect and ambipolar term can also contribute to the atmospheric induction with deviations from the winding+Ohmic balance comparable to the background field itself.
Therefore, as a second message, we stress that the GCM models which use an effective magnetic drag term should ideally incorporate the complexity of the solution that we found. Although here we have modeled only the sub-stellar point, and elsewhere the effects could be less dramatic, the main features are clear. The shear layer should host a very intense, almost purely azimuthal magnetic field with an almost purely meridional current. The Hall and ambipolar terms can induce the field in the three directions, with frequent changes of sign in all induced components.

In an upcoming work, we will use these equilibrium 1D solutions to study the 3D turbulence induced by small, forced perturbations in the shear layers, extending the ideal MHD turbulent simulations with isothermal background by \cite{soriano23} to the non-ideal regime with the more realistic $p(T)$ profiles used here.

\section*{Acknowledgments}
CSG’s work has been carried out within the framework of the doctoral program in Physics of the Universitat Autònoma de
Barcelona. CSG, DV, and AE are supported by the European Research Council (ERC) under the European Union’s Horizon 2020 research and innovation program (ERC Starting Grant ”IMAGINE” No. 948582, PI: DV). CSG, DV, and AEL acknowledge support from the “Maria de Maeztu” award to the Institut de Ciències de l’Espai (CEX2020-001058-M). We thank Carlos Palenzuela and Fabio del Sordo for the useful comments.

\section*{Data availability}
All data produced in this work will be shared on reasonable request to the corresponding author.

\input{final_version.bbl}

\appendix

\section{Effect of different boundary conditions}
\label{appendixA}

Here we present a discussion on relaxing the different boundary conditions for the $B_{x}$ component in the upper and lower limits for the simulations. The winding effect will directly depend on the magnetic field present in the domain and at the borders, i.e., the magnetic field in the inner and outer regions. For simplicity, we here restrain to the case of winding+Ohmic only. In Fig. \ref{fig:BC} top, we can see the solutions for WASP 76b} corresponding to different boundary conditions (BCs), using either a fixed value for $B_x$, or no induced current (i.e., $J_y=0$, meaning $dB_x/dz=0$).

When the induced magnetic field is fixed at $B_x=0$ G at the borders, which is the choice used in the main text, the equilibrium $B_x$ profile reaches approximately 100 G at its peak and -200 G at its minimum. Choosing a non-zero, large value (35 G in this case) of $B_{x}$ at the lower boundary (i.e., not confining the induction to the considered domain), results in a slightly larger induced field. If we instead consider cases for which the inner boundary or both boundaries have $dB_{x}/dz=0$ the magnetic field grows significantly larger and in many cases it doesn't converge: it tends to have a flat profile (i.e., current-free), but with a very large value. 

If we analyze the effect of the different $B_{x}$ boundary conditions (BCs) on the generated $Q_{j}(z)$, as shown in Fig. \ref{fig:BC} (bottom), we can observe that for high pressures ($p > 5$), there is a slight increase in $Q_{j}$ when $B_{x}$ is fixed to a certain value (purple and red lines). However, when the derivative of $B_{x}$ is set to 0, this effect is not visible. For intermediate pressures ($1 < p < 0.05$), all cases show an increase in $Q_{j}$. As discussed, this increase is proportional to the magnetic field generated, and it converges for simulations with fixed $B_{x}$ in the lower BCs, while it does not converge and continues to increase for those with the derivative of $B_{x}$ set to 0 (blue and olive lines).

Therefore, to ensure convergence and avoid an artificial unrealistic growing of the field, we choose to be conservative and we apply boundary conditions of $B_{x} = 0$ on both sides.
\begin{figure}
\centering
\includegraphics[width=.9\linewidth]{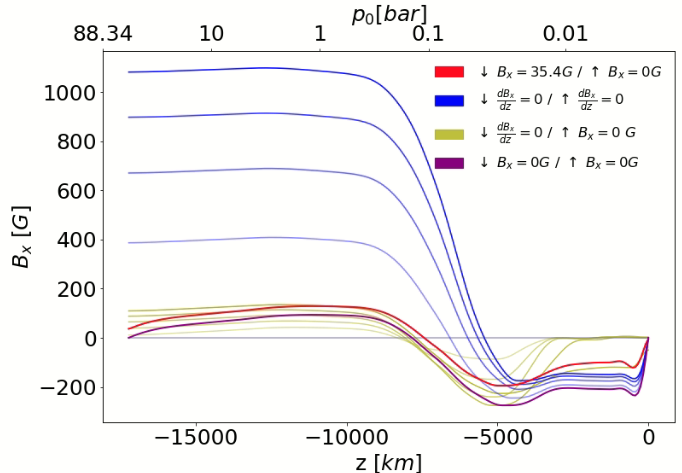}
\includegraphics[width=.9\linewidth]{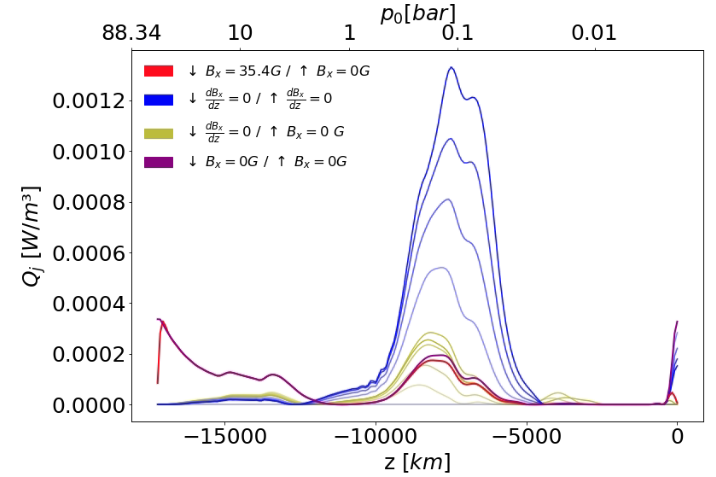}
\caption{
Comparison of the vertical profiles $B_{x} (z)$ (top) and $Q_{j} (z)$ (bottom) for WASP 76b-d3, for different BC at different times (indicated with increasingly darker shades): $B_{x}$= 35 \rm{G} in the lower border and $B_x=0$ G in the upper border (red), $dB_{x}/dz$=0 in the upper and lower border (blue), $dB_{x}/dz=0$ in the lower border and $B_{x}=0$ G in the upper border (blue) and $B_x=0$ G in the lower and upper border (purple).
}
\label{fig:BC}
\end{figure}

\section{Resolution and magnetic diffusivity}
\label{appendixB}
\begin{figure}
\centering
\includegraphics[width=.9\linewidth]{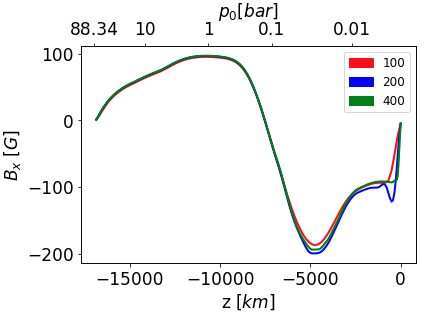}
\includegraphics[width=.9\linewidth]{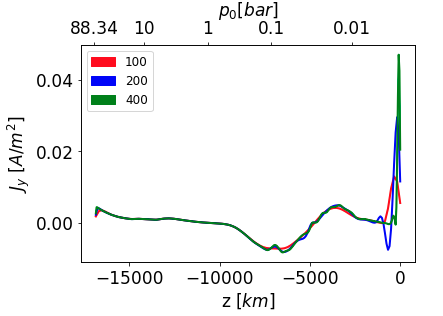}
\caption{Comparison of $B_{x}$ (top) and $J_{y}$ (bottom) at convergence for three different cases for WASP 76b: 100 (red), 200 (blue) and 400 (green) points in the vertical direction.}  
\label{fig:resolution}
\end{figure}

In this section, we aim to test that the numerical resistivity does not dominate over the analytical one. In \cite{soriano23}, the only magnetic diffusivity considered arose from numerical effects. In this paper, as discussed, the physical magnetic diffusivity has been taken into account. If we compare it with the numerical diffusivity from Appendix C in \citealt{soriano23} the physical one should dominate. Thus, a change in the resolution of the simulations should not have any effect as long as the physical magnetic diffusivity is much larger than the numerical one. In Fig. \ref{fig:resolution} we report three simulations in which the vertical resolution was changed to 100, 200, and 400 vertical points respectively.

It can be appreciated that an increase in the resolution under the same conditions does not affect the amplification of the magnetic field, since the numerical diffusivity is much smaller than the physical one, and the magnetic field throughout the domain does not practically change. If we examine the currents, $J_{y}$, the same trend can be observed. However some slightly more significant variations among the various resolutions compared to $B_{x}$ can be appreciated. This is especially so in the outermost region, where the physical resistivity drops and the numerical resistivity can be comparable to it; for this reason, the most external part of the domain, for $p \lesssim 0.01$~bar, should be considered with care. 

\end{document}